\documentclass[superscriptaddress,amsmath,amssymb,10pt,pra,floatfix,twocolumn]{revtex4-2}
\pdfoutput=1
\usepackage{graphicx}
\usepackage{color}
\usepackage{hyperref}
\usepackage{verbatim}
\usepackage{xfrac}

\begin{document}
	\title{Fine structure of the stripe phase in ring-shaped Bose-Einstein condensates with spin-orbital-angular-momentum coupling}
	\author{Y. M. Bidasyuk}
\affiliation{Physikalisch-Technische Bundesanstalt (PTB), Bundesallee 100, D-38116 Braunschweig, Germany}
	\author{K. S. Kovtunenko}
\affiliation{Department of Physics, Taras Shevchenko National University of Kyiv, Volodymyrska Str. 64/13, Kyiv 01601, Ukraine}
	\author{O. O. Prikhodko}
\affiliation{Department of Physics, Taras Shevchenko National University of Kyiv, Volodymyrska Str. 64/13, Kyiv 01601, Ukraine}
	
\begin{abstract}
	We report on a theoretical study of a ring-shaped Bose-Einstein condensate with Raman-induced spin-orbital-angular-momentum coupling.
	We analyze the structure of the ground-state of the system depending on different physical parameters and reveal a peculiar fine structure within the stripe phase on the phase diagram. 
	We demonstrate the existence of the predicted stripe sub-phases within a wide range of physical parameters, their traceability through physical observables, and compare the results with several commonly used variational approximations.
	We also show that predicted sub-phases of the stripe phase can be observed within experimentally realizable conditions. 
\end{abstract}
	
	\maketitle

\section{Introduction}

Realization of synthetic spin-orbit coupling (SOC) for cold atoms has enabled studies of many exotic quantum phenomena in a highly controllable environment of atomic Bose-Einstein condensates (BEC)
\cite{Goldman_2014,Zhai2015,ZHANG201975}.
A peculiar feature of spin-orbit-coupled systems is their energy-momentum dispersion, which contains two minima. The interplay between collisional interactions and SOC then leads to a formation of two different phases: single-momentum phase, when atoms occupy one of the two minima, and stripe phase, when both minima are occupied. In the latter case, interference of the two momentum states leads to characteristic spatially periodic density modulations, thus the name ``stripe phase''. Due to a broken translational symmetry and an emergent long-range order this phase is also often associated with the supersolidity phenomenon \cite{Li2017,RevModPhys.84.759}.

The pioneering realizations of synthetic SOC in bosonic \cite{Lin2011} and fermionic \cite{PhysRevLett.109.095301} quantum gases were based on a two-photon Raman coupling between atomic hyperfine states. 
If the Raman transition is accompanied by a change in the center-of-mass momentum of the atom, the pseudo-spin and momentum degrees of freedom become coupled in the condensate.
We will refer to such coupling as the linear SOC.
Subsequently 
a number of other techniques have been developed to achieve similar effects \cite{luo2016tunable,PhysRevLett.117.185301}. 
However, flexibility of the Raman coupling scheme together with a wide variety of accessible light fields allowed to propose and generate many different types of SOC in cold-atom setups \cite{Wu2016,PhysRevLett.118.190401}.
One specific type of Raman-induced SOC relevant for the present work is the spin-orbital-angular-momentum coupling (SOAMC), which is achieved when Raman laser beams carry a non-zero orbital angular momentum, e.g. Laguerre-Gaussian (LG) beams \cite{PhysRevLett.121.113204,PhysRevLett.122.110402,PhysRevA.93.013629,PhysRevA.94.033627,PhysRevA.91.063627,PhysRevResearch.2.033152,Chiu_2020}.

Interestingly, most of the previous studies focus on the SOAM coupling in harmonically trapped condensates. In this case the phase singularity of the LG Raman beam is located inside the condensate, which makes the system considerably different from the one with linear SOC. Most notably, such a singularity may lead to a formation of various vortex states in the ground state \cite{PhysRevA.91.033630,PhysRevA.92.033615,PhysRevA.102.063328}.

In the present work we investigate ground-state phases of a toroidal SOAM-coupled atomic condensate. 
If the trap potential has a toroidal shape, then the phase singularity is excluded from the condensate region.
This leads to absence of vortices in the ground state and makes analogies with linear SOC more straightforward \cite{PhysRevA.91.063627}.
Using an accurate numerical minimization of the mean-field energy as well as several simplified variational approximations we reveal and analyze a peculiar internal structure of the stripe phase.
The predicted sub-phases produce distinctive density distributions of the ground state and can be detected through certain physical observables, such as the average angular momentum of the system.

The paper is organized as follows.
In Sec. II we introduce our model physical system and several commonly used approximations, which are necessary to reduce the Hamiltonian to a convenient one-dimensional form.
In Sec. III we analyze the spectrum of the single-particle Hamiltonian and establish the phase diagram of the noninteracting system. 
In Sec. IV we present several variational models extending on the single-particle solutions to account for the effects of interactions.
We derive a number of analytical relations to estimate the region in parameter space, where the stripe phase may exist.
In Sec. V we discuss the phase diagram of the interacting system and structure of the ground state within different phases. We reveal and analyze the internal fine structure of the stripe phase and highlight the apparent limitations of the variational approximations.
In Sec. VI we show example solutions for a realistic three-dimensional system and confirm that previously identified features of the stripe phase can be observed in a realistic scenario.
Finally, Sec. VII contains a short summary and conclusions of the work.
	
	\section{Model equations}\label{sec:model}

We consider a zero-temperature atomic BEC in a toroidal trap.
Spin-orbit coupling in the condensate is achieved by applying a continuous coherent Raman coupling between two different hyperfine substates from the ground state manifold \cite{Lin2011}. 
In this case we can effectively use spin-1/2 algebra and characterize the condensate by a (pseudo)spinor field
	\begin{equation}\label{eq:psi2c}
	\Psi = \begin{bmatrix}
	\psi_a(\mathbf{r}) \\ \psi_b(\mathbf{r})
	\end{bmatrix}.
	\end{equation}
The single-particle Hamiltonian is then represented as a $2\times2$ matrix
\begin{equation}\label{eq:ham0}
\mathcal{H}_0 = \frac{\hbar^2\nabla^2}{2M}\mathbb{I}_2 + V (\mathbf{r}) \mathbb{I}_2 + \mathcal{H}_\mathrm{c},
\end{equation}
consisting of (spin-independent) kinetic energy and trap potential, as well as the spin-coupling term $\mathcal{H}_\mathrm{c}$.
In cylindrical coordinates $\mathbf{r} = (r,\varphi,z)$ toroidal trap can be approximated by harmonic potentials in radial and longitudinal directions
\begin{equation}\label{eq:torpot}
V(\mathbf{r}) = \frac{M}{2}\left[\omega_r^2 (r-r_0)^2 + \omega_z^2 z^2 \right],
\end{equation}
where $\omega_z$ and $\omega_r$ are longitudinal and radial trap frequencies, respectively.
The Raman coupling term has the following form within the rotating wave approximation:
\begin{equation}\label{eq:2praman}
\mathcal{H}_c = \frac{\hbar}{2}\begin{bmatrix} -\tilde\delta & \tilde\Omega f(r) e^{-2im_0\varphi} \\ \tilde\Omega f(r) e^{2im_0\varphi} & \tilde\delta \end{bmatrix}.
\end{equation}
Such form of the coupling Hamiltonian arises when Raman transition is induced by two copropagating Laguerre-Gaussian (LG) beams with orbital angular momenta $m_0$ and $-m_0$ and with their beam axes aligned along the $z$-axis \cite{PhysRevLett.122.110402,PhysRevA.91.063627}. 
In this case the Raman absorption and stimulated emission sequence transfers $2m_0$ angular momentum quanta to the condensate. Previous experiments with SOAM coupling in BEC were conducted mostly with $m_0=1$. However, it is also possible to generate and manipulate LG laser modes with much higher OAM \cite{Campbell:12,Fickler13642}. Throughout the present work we use the value $m_0=10$, which is large enough to see a more general picture, valid for any OAM transfer implemented within the Raman setup.

Other quantities that enter in the coupling Hamiltonian (\ref{eq:2praman}) are two-photon frequency detuning $\tilde\delta$ and the Raman coupling with the amplitude $\tilde\Omega$ and the radial distribution $f(r)$ reflecting the intensity profile of the LG laser beam. If the BEC ring is thin and its radius $r_0$ coincides with the intensity maximum of the Raman coupling, then the specific shape of $f(r)$ is of minor importance \cite{PhysRevA.91.063627,PhysRevA.96.011603}.

In order to make the system accessible for analytical investigation we can use several commonly applied transformations and approximations.
First, we consider the parametric regime when characteristic energy scales for atoms movement in $r$ and $z$ dimensions are much higher then those of the Raman coupling, i.e. $\omega_r,\omega_z \gg \tilde \Omega,\tilde \delta$.
This means that movement in $r$ and $z$ directions can be considered frozen and these dimensions can be factorized in the wave function.
We also introduce a unitary transformation of the wave function in the angular dimension, such that
\begin{equation}\label{eq:utrans}
\begin{array}{l}
\displaystyle \psi_a(\mathbf{r}) \rightarrow \psi_a(\varphi) \chi(r,z)  e^{-im_0\varphi},\\[2mm]  
\displaystyle \psi_b(\mathbf{r}) \rightarrow \psi_b(\varphi) \chi(r,z) e^{im_0\varphi},
\end{array}
\end{equation}
with normalization conditions
\begin{equation}\label{eq:rz-norm}
\int dr dz |\chi(r,z)|^2=1,\quad \int dr dz f(r)|\chi(r,z)|^2=1. 
\end{equation}
The above transformation removes explicit dependence of the Hamiltonian on the angular coordinate $\varphi$ and allows to integrate out $r$ and $z$ dimensions,
leaving us with the following one-dimensional Hamiltonian

\begin{equation}\label{eq:h0dim}
\mathcal{\tilde H}_{\varphi} = \frac{ \hbar^2}{2Mr_0^2} \left(i\partial_{\varphi}+m_0 \sigma_z\right)^2 + \frac{\hbar\tilde \Omega}2  \sigma_x - \frac{\hbar\tilde \delta}2 \sigma_z,
\end{equation}
where  $\sigma_z$ and $\sigma_x$ are the standard $2 \times 2$ Pauli matrices. 
The Hamiltonian can be further rewritten in the dimensionless form using the energy unit $\epsilon=\hbar^2/(Mr_0^2)$, which is a characteristic energy scale of rotation in the system \cite{PhysRevA.87.013619}. We finally get the following dimensionless Hamiltonian:

\begin{equation}\label{eq:h0}
\mathcal{H}_{\varphi} = \frac{ (L_z-m_0 \sigma_z)^2}{2} + \frac{\Omega}2 \sigma_x - \frac{\delta}2 \sigma_z,
\end{equation}
where $\Omega=\hbar\tilde\Omega/\epsilon$, $\delta=\hbar\tilde\delta/\epsilon$, $L_z=-i \partial_\varphi$ are dimensionless Raman coupling strength, detuning and the angular momentum operator respectively. A single-particle Hamiltonian of this general form is commonly used to study spin-orbital coupling in two-component spinor BECs \cite{Lin2011,PhysRevLett.108.225301,PhysRevA.91.063627}. 

The ground state of a weakly interacting spinor condensate is defined through a minimization of the Gross-Pitaevskii energy functional
\begin{multline}\label{eq:gpe_energy1d}
\!\!\!E = \int\limits_0^{2\pi} d\varphi \left[  
\begin{pmatrix}
\psi_a^* &
\psi_b^*
\end{pmatrix}
\mathcal{H}_{\varphi} 
\begin{pmatrix}
\psi_a\\
\psi_b
\end{pmatrix}  + \frac{g_{aa}}2 |\psi_a|^4  + \frac{g_{bb}}2 |\psi_b|^4 \right. \\ \left. + g_{ab} |\psi_a|^2 |\psi_b|^2
\right].
\end{multline}
where $g_{ij}$ are (one-dimensional) nonlinear interaction constants between atoms in states $i$ and $j$, also expressed in the units of $\epsilon$. We assume here for simplicity that interactions are spin-symmetric ($g_{aa}=g_{bb}=g$) and set the normalization condition as
\[
\int\limits_0^{2\pi} d\varphi \left(|\psi_a|^2 + |\psi_b|^2\right) = 1,
\]
so that $g$ and $g_{ab}$ are proportional to the total number of atoms in the condensate. Additionally, we limit the present study to the situations when interactions are repulsive and two components of the condensate are fully miscible, i.e. $g>g_{ab}>0$.

Before we continue with analyzing the spectrum and eigenstates of the Hamiltonian (\ref{eq:h0}) it is worth briefly discussing its symmetry properties. 
The most important is the rotational symmetry, stemming from commutation of the Hamiltonian (\ref{eq:h0}) with the angular momentum operator $L_z$. This symmetry plays here the same role as the translational symmetry in the systems with linear SOC. Spontaneous breaking of rotational symmetry leads to formation of a stripe phase which is one of the most stunning features of the spin-orbit coupled systems.
One crucial difference between linear SOC and SOAMC, however, arises from the periodic boundary conditions producing a quantization of (angular) momentum, which is absent in systems with linear SOC. As we will see in the next sections, this difference leads to a considerably richer phase diagram of the SOAM-coupled ring system in comparison to a thoroughly studied uniform BEC with linear-momentum SOC.

Another important symmetry of the Hamiltonian (\ref{eq:h0}), which exists only with $\delta=0$, is a $\mathbb{Z}_2$ symmetry often associated with the time reversal. The corresponding symmetry transformation combines the complex conjugation of the wave function and spin inversion:
\begin{equation}\label{eq:t-op}
\mathcal{T}\begin{bmatrix}
\psi_a \\ \psi_b
\end{bmatrix} = \begin{bmatrix}
\psi_b^* \\ \psi_a^*
\end{bmatrix}.
\end{equation}
Time reversal symmetry also can be spontaneously broken in the ground state leading to formation of polarized phases \cite{PhysRevA.102.063328}. Any non-zero detuning $\delta$ leads to the absence of such symmetry in the Hamiltonian. Still, the symmetry transformation (\ref{eq:t-op}) plays an important role for the eigenstates of the system, as will be seen in the next section.

\section{Noninteracting system}

\begin{figure}[tbp]
	\centering
	\includegraphics[width=\linewidth]{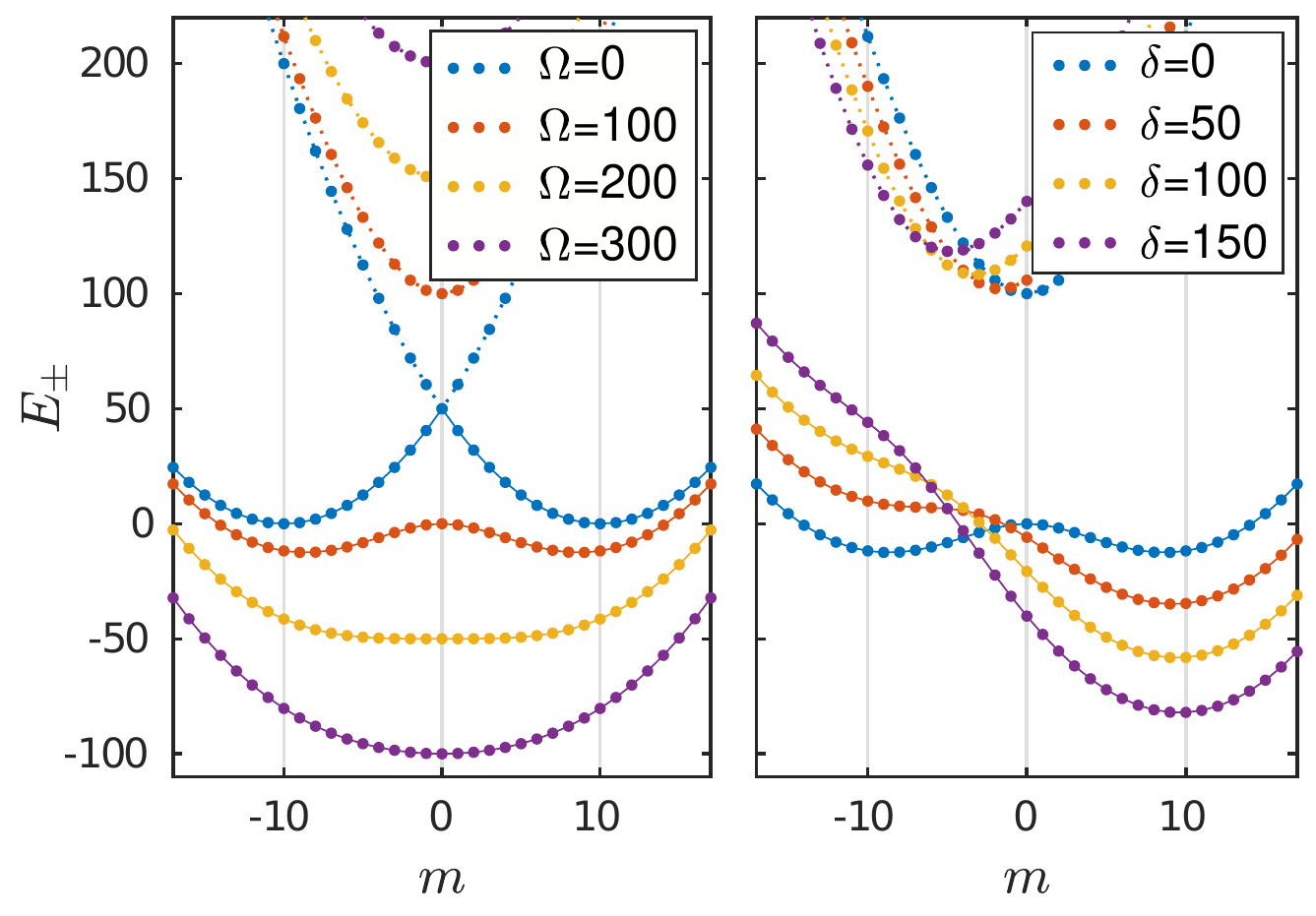}
	\caption{Single-particle energy spectrum defined by Eq.~(\ref{eq:sp_spectr}) with $m_0=10$. 
	Left panel corresponds to $\delta=0$ and various values of $\Omega$.
	Right panel corresponds to $\Omega=100$ and various values of $\delta$. 
	Lines are guides for the eye marking the lower (solid lines) and upper (dotted lines) branches of the spectrum.}
	\label{fig:sp_spectr}
\end{figure}

As a result of the rotational symmetry of the Hamiltonian (\ref{eq:h0}), its 
eigenstates can be characterized by a well defined angular momentum projection.
The corresponding energy spectrum can then be written in terms of the angular quantum number $m$ as follows
\begin{equation}\label{eq:sp_spectr}
E_{\pm} (m) = \frac12 \left[m_0^2+m^2 \pm \sqrt{\Omega^2+(2m_0m+\delta)^2}\right].
\end{equation}
The energy spectrum consists of two branches (see Fig.~\ref{fig:sp_spectr}). The lower branch $E_-$ may contain two minima. Depending on $\Omega$ and $\delta$ these minima may be located at different values of $m$ within the range $[-m_0,m_0]$ and either one or both of them correspond to the ground state of the system. Understanding which value of $m$ represents the ground state will be the main goal of the present section. 

\begin{figure}[tbp]
	\centering
	\includegraphics[width=\linewidth]{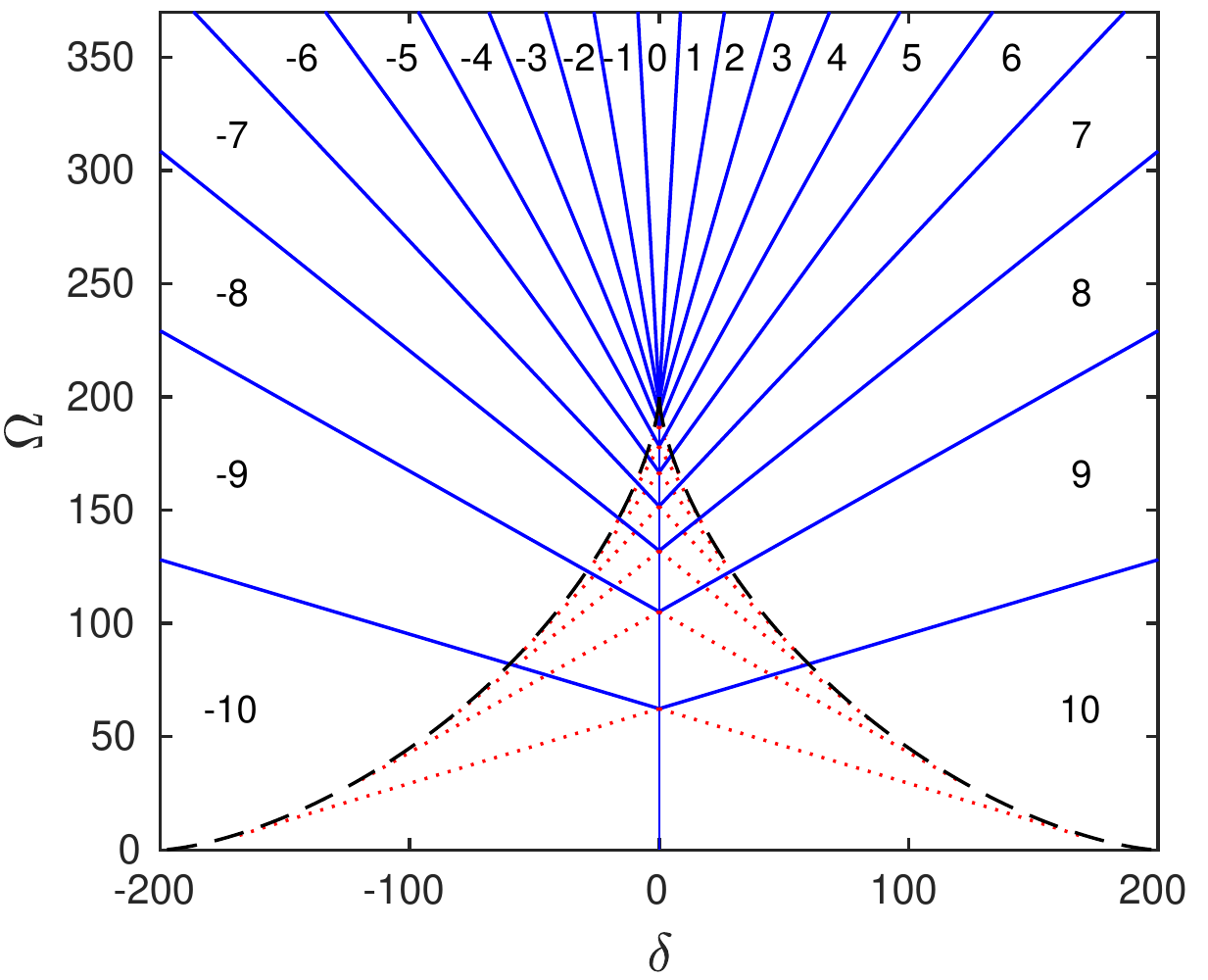}
	\caption{Single-particle ``phase diagram''  for $m_0=10$. Blue lines and numbers between $-10$ and $10$ show the regions where corresponding value of $m$ represents the ground state of the system. Red dotted lines show the same for the second minimum. Black dashed line shows the region where two minima exist in the energy spectrum and is defined by Eq.~(\ref{eq:cusp})}
	\label{fig:sp_pd}
\end{figure}

If $m$ was a continuous variable, we would find the ground state from the extremum condition $\partial E_-/\partial m = 0$. This condition, when solved with respect to $\Omega$, yields the following expression:
\begin{equation}\label{eq:omega_sp_cont}
\Omega= \left(2m_0 +\frac{\delta}{m}\right) \sqrt{m_0^2-m^2}.
\end{equation}
However, as $m$ is a discrete quantum number, we have to use a different approach. We can find regions in $(\Omega,\delta)$ space where specific value $m$ correspond to the ground state by solving the equation $E_-(m)=E_-(m+1)$. 
This equation can be solved for $\Omega$ analytically, providing a critical coupling, where the ground states switches from $m$ to $m+1$:
\begin{equation}\label{eq:omega_sp}
\Omega_{m} = \sqrt{\left[\left(2m_0 +\frac{\delta}{m+1/2}\right)^2 - 1 \right]\left[m_0^2- \left(m+\frac12\right)^2\right]}.
\end{equation}
This expression allows us to define the regions in $(\Omega,\delta)$-space where angular quantum number $m$ represents the ground state, providing effectively a single-particle ``phase diagram'' of the system, which is shown in Fig.~\ref{fig:sp_pd}. Similar expression  can be found in Ref.~\cite{PhysRevA.91.063627}. 
Interestingly, if we consider $(2m_0)^2 \gg 1$ then Eq.~(\ref{eq:omega_sp}) simplifies to
\begin{equation}\label{eq:omega_sp_cont2}
\Omega_{m}\approx \left(2m_0 +\frac{\delta}{m+1/2}\right) \sqrt{m_0^2-\left(m+\frac12\right)^2},
\end{equation}
which is equivalent to Eq.~(\ref{eq:omega_sp_cont}) evaluated at the half-integer value $m+1/2$. Such equivalence between the continuous and discrete extremum conditions will be especially useful in the next section for the analysis of the interacting system.

If the energy $E_-(m)$ contain two minima, then the extremum conditions derived above will be satisfied not only for the ground state, but also for the second energy minimum and the maximum in between. 
Therefore, we can also find regions in parameter space where certain quantum number $m$ corresponds to a metastable excited state as well as the region where two energy minima exist (see Fig.~\ref{fig:sp_pd}).
This region, where the energy spectrum has two minima, is localized at small values of $\Omega$ and $\delta$ and bounded by the astroid curve 
\begin{equation}\label{eq:cusp}
\Omega^{2/3} + |\delta|^{2/3} = \Big(2m_0^2\Big)^{2/3}.
\end{equation}

It will be useful for the discussions in the next section to also
introduce the eigenstates corresponding to the energy spectrum (\ref{eq:sp_spectr}). These normalized eigenstates read

\begin{equation}\label{eq:sp_wavefunc}
\Psi_m = \begin{bmatrix}
\psi_a(\varphi)\\
\psi_b(\varphi)
\end{bmatrix}
=
\frac{1}{\sqrt{2\pi}}
\begin{bmatrix}
\cos(\theta/2)\\
-\sin(\theta/2)
\end{bmatrix} e^{i m \varphi},
\end{equation}
with 
\[
\theta = \arctan\left(\frac{\Omega}{2m_0m+\delta}\right).
\]
Eigenstates in such a form describe both branches of energy spectrum. The values $0<\theta<\pi$ correspond to the lower energy branch $E_-$, while $\pi<\theta<2\pi$ corresponds to $E_+$.
Eigenstates with the opposite angular quantum numbers $m$ and $-m$ are related through a time-reversal transformation (\ref{eq:t-op}):
\begin{equation}\label{eq:trsym}
\Psi_{-m} = \mathcal{T} \Psi_m = \frac{1}{\sqrt{2\pi}}
\begin{bmatrix}
-\sin(\theta/2)\\
\cos(\theta/2)
\end{bmatrix} e^{-i m \varphi}.
\end{equation}
If $\delta = 0$ then the states $\Psi_m$ and $\Psi_{-m}$ have the same energy due to a time-reversal symmetry of the Hamiltonian.

On a final note, the energy spectrum (\ref{eq:sp_spectr}) is symmetric with respect to $\Omega$, and also symmetric with respect to $\delta$ with a replacement $m\rightarrow-m$. We can therefore restrict all further analysis to $\Omega>0$ and $\delta>0$ without a loss of generality.

\section{Variational treatment of the interacting system}

We now extend the above one-dimensional analysis to the case of interacting system. 
In the presence of interactions minimization of the energy functional (\ref{eq:gpe_energy1d}) can not be performed analytically. 
Nevertheless, some analytical treatment is possible within the variational approach.
But first, for the convenience of further analysis we introduce instead of nonlinear parameters $g$ and $g_{ab}$ new parameters $G_1 = (g+g_{ab})/8\pi$ and $G_2 = (g-g_{ab})/8\pi$. The energy functional (\ref{eq:gpe_energy1d}) is then rewritten in an equivalent form

\begin{multline}\label{eq:gpe_energy2}
\!\!\!E = \int\limits_0^{2\pi} d\varphi \left[  
\begin{pmatrix}
\psi_a^* &
\psi_b^*
\end{pmatrix}
\mathcal{H}_{\varphi} 
\begin{pmatrix}
\psi_a\\
\psi_b
\end{pmatrix}  + 2\pi G_1 \left(|\psi_a|^2 + |\psi_b|^2\right)^2 \right. \\ \left. + 
2\pi G_2 \left(|\psi_a|^2 - |\psi_b|^2\right)^2
\right].
\end{multline}
It is worth mentioning here that for the most experimental realizations of spin-orbit-coupled BECs with alkali atoms the nonlinear interaction coefficients satisfy the inequality $0<G_2\ll G_1$ \cite{PhysRevLett.116.160402}.

\subsection{Single-mode ansatz}

The simplest, yet quite instructive approach is to take a single noninteracting solution $\Psi_m$ from Eq.~(\ref{eq:sp_wavefunc}) as the variational trial function and consider $m$ and $\theta$ as variational parameters. After inserting (\ref{eq:sp_wavefunc}) into the energy functional (\ref{eq:gpe_energy2}) we get
\begin{multline}
E(m,\theta) = \frac{m_0^2+m^2}{2} - \left(m_0 m + \frac\delta2\right)\cos\theta \\ - \frac\Omega2\sin\theta  + G_1 + G_2\cos^2\theta. 
\end{multline}
This expression can be minimized analytically with respect to $m$ and $\theta$ only if $m$ is considered as a continuous variable. We therefore follow the analogy with the noninteracting case and express the critical coupling $\Omega_m$ for the transition between $m$ and $m+1$ ground states by inserting $m+1/2$ into a solution of the system of equations $\partial E/\partial m=0$ and $\partial E/\partial \theta=0$. We obtain the following expression very similar to Eq.~(\ref{eq:omega_sp_cont2}):
\begin{equation}\label{eq:omega_sp1}
\Omega_{m} = \left[ 2 m_0 +\frac{\delta}{m+1/2}  - \frac{4G_2}{m_0} \right] \sqrt{m_0^2- \left(m+\frac12\right)^2}.
\end{equation}
We see that within the single mode ansatz interactions lead only to shifts of boundaries on the single-particle phase diagram. 
Qualitatively, the picture remains the same as in Fig.~\ref{fig:sp_pd}.  
Validity of the above expression will be evaluated numerically in the next section.
It is important to note that Eq.~(\ref{eq:omega_sp1}) does not depend on $G_1$ and in the case of spin-independent collisional interactions, i.e. $g=g_{ab}$, and consequently $G_2=0$, it becomes fully identical to the result of noninteracting system (\ref{eq:omega_sp_cont2}).

\subsection{Two-mode ansatz}

As already mentioned above, the stripe phase is associated with simultaneous occupation of two minima of the energy spectrum.
Therefore, the most commonly used variational ansatz  that is able to describe the stripe phase is based on a linear combination of two wave functions with opposite angular momenta $m$ and $-m$ (see e.g. \cite{PhysRevLett.108.225301,PhysRevA.91.063627,PhysRevResearch.2.033152,Chiu_2020}):
\begin{equation}\label{eq:2wave}
\Psi = C_+ \Psi_m + C_- \Psi_{-m}.
\end{equation}
The functions $\Psi_m$ and $\Psi_{-m}$ are defined by Eq.~(\ref{eq:sp_wavefunc}) and related through a time-reversal transformation (\ref{eq:trsym}). The coefficients $C_+$ and $C_-$ are constrained by a normalization condition $|C_+|^2+|C_-|^2=1$. Consequently, we have a variational problem with three independent variational parameters: $m$, $\theta$ and $\beta = |C_+|^2-|C_-|^2$.
Without a loss of generality we may consider $|C_+| \geq |C_-|$ and restrict $\beta$ within the range $[0,1]$.
After inserting (\ref{eq:2wave}) into (\ref{eq:gpe_energy2}) we get the following expression of energy as a function of variational parameters:
\begin{multline}\label{eq:en-2wave}
E(m,\theta,\beta) = \frac{m_0^2+m^2}{2} - \left(m_0 m + \frac{\beta \delta}2\right)\cos\theta \\ - \frac\Omega2\sin\theta  + G_1 + G_1\frac{1-\beta^2}{2}\sin^2\theta  + G_2 \beta^2 \cos^2\theta.
\end{multline}
Many other useful physical quantities can also be readily expressed in terms of our variational parameters. These include spin polarizations
\begin{equation}
\langle \sigma_z \rangle = \beta \cos \theta, \quad \langle \sigma_x \rangle = - \sin \theta,
\end{equation}
expectation value of angular momentum
\begin{equation}\label{eq:lz-an}
\langle L_z \rangle = m \beta,
\end{equation}
and its standard deviation
\begin{equation}
\Delta L_z=\sqrt{\langle L_z^2 \rangle-\langle L_z \rangle^2} = m \sqrt{1-\beta^2},
\end{equation}
as well as the particle density distribution
\begin{multline}\label{eq:dens-2m}
|\Psi(\varphi)|^2 = |\psi_a|^2+|\psi_b|^2 \\ = \frac{1}{2\pi} \left[1+\sqrt{1-\beta^2} \sin \theta \cos (2m\varphi)\right].
\end{multline}
The variational parameter $\beta$ describes the degree of mixing between the two angular momentum modes. With $\beta=1$ we get a state with well defined angular momentum and recover the above single-mode result. It corresponds to the single-momentum (SM) phase of the system. With $\beta < 1$ two angular momentum components are populated. Such states contain $2m$ oscillations in the particle density and represent the stripe phase of the system. 

In general, the energy functional (\ref{eq:en-2wave}) can be minimized only numerically.
There are, however, several important particular cases that can be studied analytically in order to estimate the region in $(\Omega,\delta)$ space where the stripe phase may exist.

First, Eq.~(\ref{eq:en-2wave}) can be relatively easily analyzed in the limit $\Omega \rightarrow 0$. In this case the energy minimum is achieved with
\begin{equation}\label{eq:small-delta}
m=m_0,\quad \theta=0, \quad \beta=\frac{\delta}{4G_2}
\end{equation}
if $\delta<\delta_\mathrm{c} = 4 G_2$.
This region corresponds to the stripe phase. If $\delta>\delta_\mathrm{c}$ then the energy is minimized by
\begin{equation}
m=m_0,\quad \theta=0, \quad \beta=1,
\end{equation}
representing the SM phase with the momentum $m_0$.

Another important limit case is zero detuning $\delta=0$.
With $\delta=0$ the energy (\ref{eq:en-2wave}) becomes a linear function of $\beta^2$. 
Therefore, energy minimum with respect to $\beta$ can be achieved only at one of the limit values. 
The value $\beta=1$ represents the SM phase, while $\beta=0$ corresponds to the stripe phase. The boundary between the two phases can be found by minimizing separately $E(m,\theta,0)$ and $E(m,\theta,1)$ and finding the crossing between the two. Minimization of the single-momentum energy $E(m,\theta,1)$ was already discussed above. Minimization of the stripe-phase energy $E(m,\theta,0)$ is more complicated, since in this case application of continuous extremum condition on $m$ is not justified. Relatively simple solutions are possible to obtain in strongly-interacting regime, when $G_1 \gg \Omega$. In this case the energy minimum is achieved with the following values of variational parameters
\begin{equation}
m=m_0,\quad \sin\theta \approx \frac{\Omega}{2(G_1+m_0^2)}, \quad \beta=0.
\end{equation}
For the transition between the stripe phase and the SM phase there are two possible distinct cases. If 
\begin{equation}\label{eq:g2-cond}
G_2>\frac{1}{2}\left[ \sqrt{G_1(m_0^2 + G_1)}-G_1\right],
\end{equation}
then the transition is observed at the critical coupling strength 
\begin{equation}\label{eq:omc-2mode1}
\Omega_{\mathrm{c}} = 2\left[ m_0^2 + G_1 - \sqrt{G_1(m_0^2 + G_1)} \right].
\end{equation}
Below this critical coupling ($\Omega<\Omega_{\mathrm{c}}$) the energy minimum corresponds to the stripe phase and if $\Omega>\Omega_{\mathrm{c}}$, the energy is minimized with $m=0$ and $\beta=1$ corresponding to a SM phase with zero angular momentum.

If the condition (\ref{eq:g2-cond}) is not satisfied, then the transition occurs to the state with non-zero angular momentum. In this case, analogously to the above single-mode discussion, the SM energy $E(m,\theta,1)$ needs to be minimized approximating $m$ as a continuous variable. The resulting critical coupling reads
\begin{equation}\label{eq:omc-2mode2}
\Omega_{\mathrm{c}} = 2\sqrt{\frac{2G_2(m_0^2 + G_1)(m_0^2 - 2G_2)}{G_1+2G_2}}.
\end{equation}
Eqs. (\ref{eq:g2-cond}) -- (\ref{eq:omc-2mode2}) allow us to estimate the range of Raman couplings $\Omega$ where the stripe phase may exist.
These results are consistent with previous findings for systems with linear SOC \cite{PhysRevLett.108.225301}.

\subsection{Multi-mode ansatz}

In order to develop a more accurate representation for possible ground states, which would generalize the above two-mode ansatz, we start with writing the wave function as a following superposition of angular modes:
\begin{equation}\label{eq:nwave}
\Psi = \sum\limits_{n=-\infty}^{\infty} 
\begin{pmatrix}
A_n \\ B_n
\end{pmatrix} 
e^{i(m-n k)\phi},
\end{equation}
with $k > m \geq 0$, so that $m$ represents the lowest non-negative angular momentum component and $m-k$ is the lowest negative one.
The idea behind this representation is that two central components $n=0$ and $n=1$ with angular momenta $m$ and $m-k$ represent simultaneous population of two minima in the energy spectrum. All other higher order components appear due to nonlinear mixing caused by interactions.
Similar representations were used previously to describe the stripe phase in a system with linear Raman-induced SOC \cite{PhysRevLett.110.235302} and SOAMC \cite{Chiu_2020}. 

Importantly, the infinite sum (\ref{eq:nwave}) is still an exact representation of both stripe and SM states of the system.
In the latter case the sum contains only one pair of non-zero  coefficients $A_0$ and $B_0$.
To use the representation (\ref{eq:nwave}) as a variational trial function, the sum must be truncated to a finite number of terms. Truncating the the sum to only two central components $n=0,1$ can be considered as a generalization of the two-wave ansatz (\ref{eq:2wave}). 
The variational problem then contains in total five independent variational parameters. Those are $m$, $k$, and three out of the four coefficients $A_0$, $B_0$, $A_1$, $B_1$, which are constrained by the normalization condition. 

In order to see the role of higher-order angular modes
for the structure of the stripe phase 
we will also use four central components of Eq.~(\ref{eq:nwave})
as a trial function, such that
\begin{multline}\label{eq:4mode}
\Psi =  
\begin{pmatrix}
A_{-1} \\ B_{-1}
\end{pmatrix} 
e^{i (m+k) \phi}+
\begin{pmatrix}
A_0 \\ B_0
\end{pmatrix} 
e^{i m \phi} \\ +
\begin{pmatrix}
A_1 \\ B_1
\end{pmatrix} 
e^{i (m-k) \phi}+
\begin{pmatrix}
A_{2} \\ B_{2}
\end{pmatrix} 
e^{i (m-2k) \phi}.
\end{multline}
The coefficients $A_n$ and $B_n$ and characteristic wave numbers $m$ and $k$ are considered as variational parameters.
Accounting for the normalization condition, we have then in total 9 independent variational parameters to be optimized.
A similar variational approximation was previously used in Ref.~\cite{Chiu_2020} showing a considerable improvement over a two-mode atsatz for the determination of a stripe contrast.

\section{Phase diagram}

We now use the variational ans\"atze described above as well as the full numerical minimization of the energy functional (\ref{eq:gpe_energy2}) to build and analyze a phase diagram of the system. 
We have in total four external parameters that can be used to drive the phase transitions. Those are Raman coupling strength $\Omega$, detuning $\delta$, and nonliear interaction coefficients $G_1$ and $G_2$. Phase diagrams can be built by fixing two of these parameters and tracing the ground state of the system through a range of values for other two.

\begin{figure}[tbp]
	\centering
	\includegraphics[width=\linewidth]{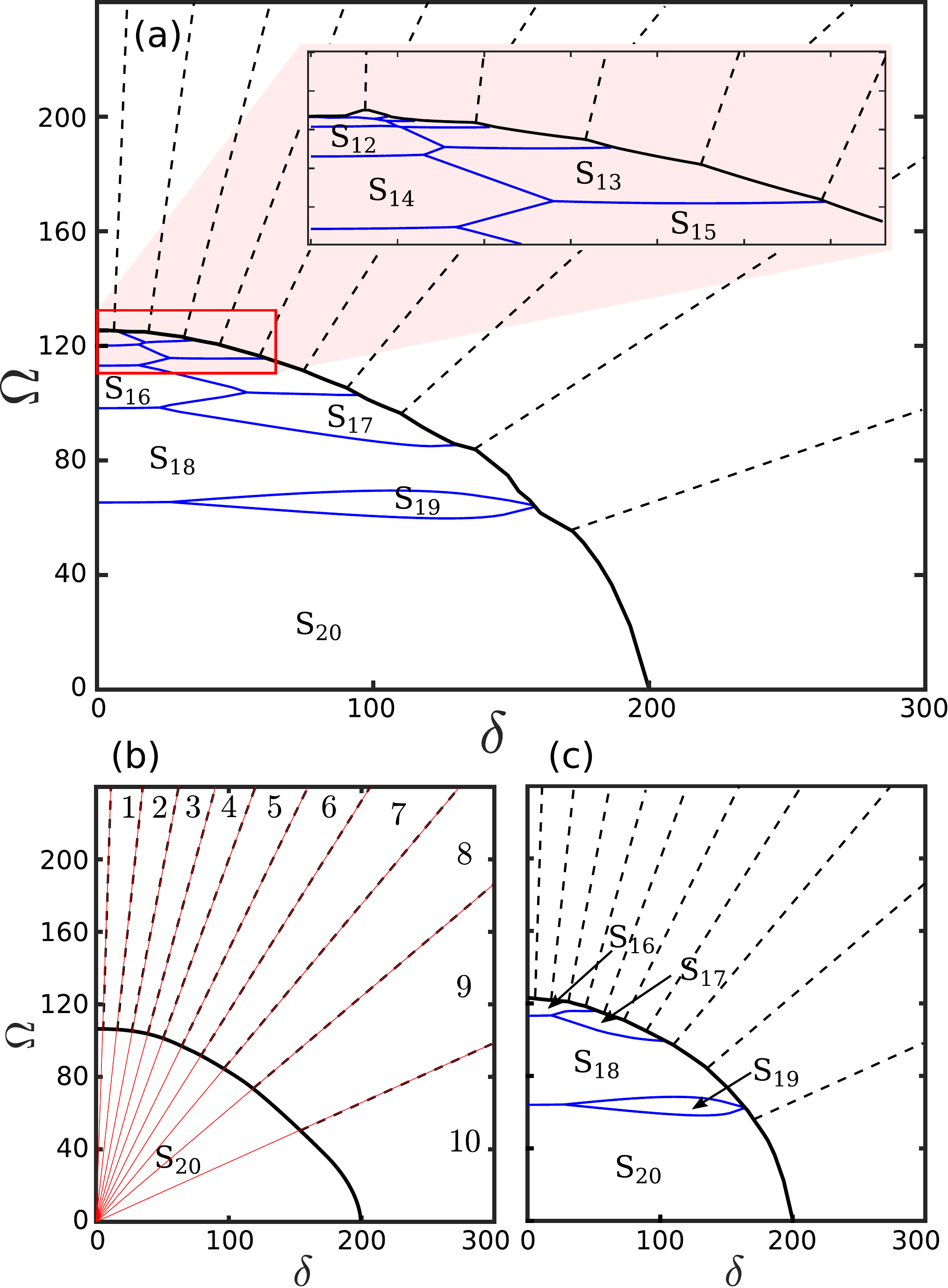}
	\caption{Phase diagram of SOAM-coupled ring system with $m_0=10$, $G_1=500$, $G_2=50$ calculated with full numerical minimization (a), two-mode variational ansatz (b) and four-mode atsatz (c).
	Thick black line shows the boundary between the stripe phase and the single-momentum phase. Dashed black lines show boundaries between ground states with different well-defined angular momentum [marked also with numbers on panel (b)]. Blue lines mark boundaries between different types of stripe states.
	Thin red lines on panel (b) represent the result of Eq.~(\ref{eq:omega_sp1}).
	}
	\label{fig:int_pd}
\end{figure}

Fig.~\ref{fig:int_pd} shows the calculated phase diagram in $(\Omega,\delta)$-plane for fixed values of $G_1=500$ and $G_2=50$.
The phase diagram consists of two large regions representing two phases of the system. The stripe phase is observed in the the region of small values of $\Omega$ and $\delta$ and single-momentum (SM) phase covers the rest of the parameter space. 

Both phases exhibit certain internal structure.
Inside the SM phase we observe regions of different well-defined angular momentum. Boundaries of these regions are similar to the single-particle phase diagram and can be very accurately approximated by Eq.~(\ref{eq:omega_sp1}). The observed internal structure of the stripe phase is much more peculiar, since it does not directly follow from any of the above analytical predictions and was not observed in previous studies.
The stripe phase consists of a number of sub-phases, which we label as S$_k$. The index $k$, introduced here to distinguish these sub-phases, represents the number of periodic density modulations (stripes) observed in the particle density of the ground state. This number is always integer due to the periodicity of the wave function.
In the region of small $\delta$ and close to the transition to the SM phase we observe an increasing number of stripe sub-phases, which become difficult to resolve numerically.

\begin{figure}[tbp]
	\centering
	\includegraphics[width=\linewidth]{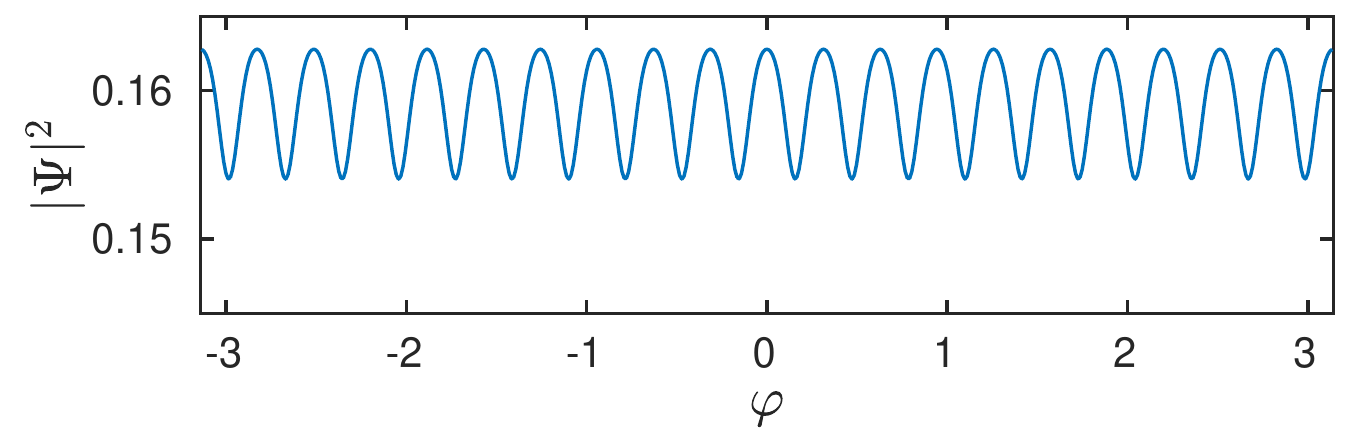}
	\includegraphics[width=\linewidth]{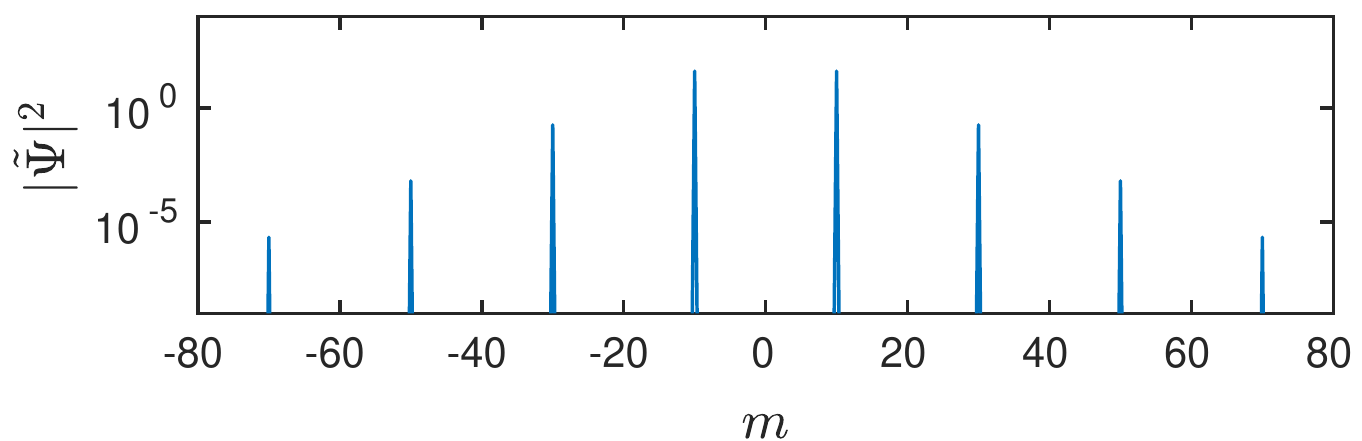}
	\caption{
	Numerically calculated particle density (top panel) and power spectrum (bottom panel) of the ground state at $\Omega=60$, $\delta=0$. 
	The state corresponds to the sub-phase $S_{20}$, as marked in Fig.~\ref{fig:int_pd}, and contains 20 density modulations. Four central peaks in $|\widetilde \Psi|^2$ are located at $ m= \{\pm 10, \pm 30 \}$. Note the logarithmic scale of the vertical axis on the lower panel.}
	\label{fig:sol1}
\end{figure}

\begin{figure}[tbp]
	\centering
	\includegraphics[width=\linewidth]{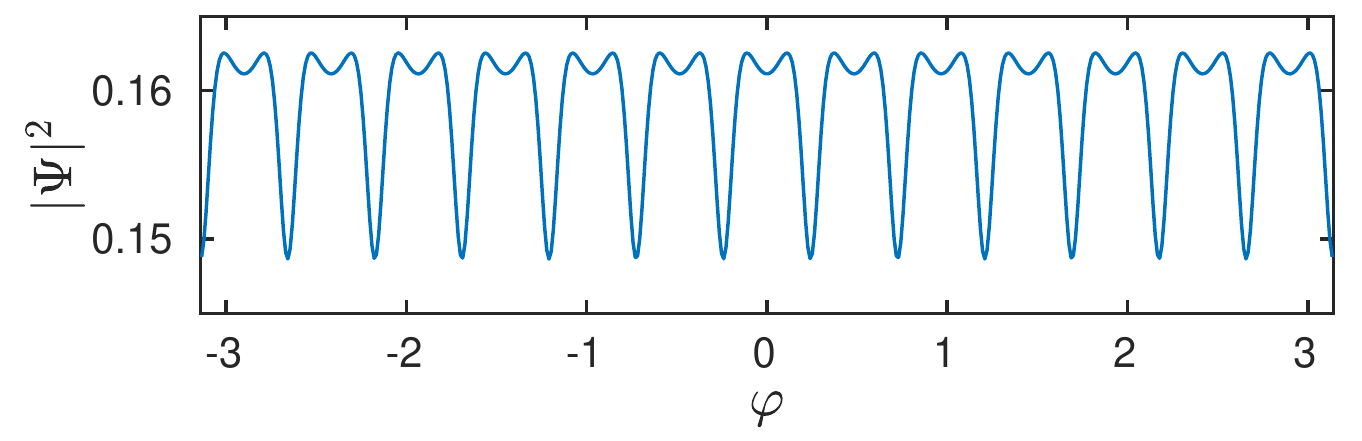}
	\includegraphics[width=\linewidth]{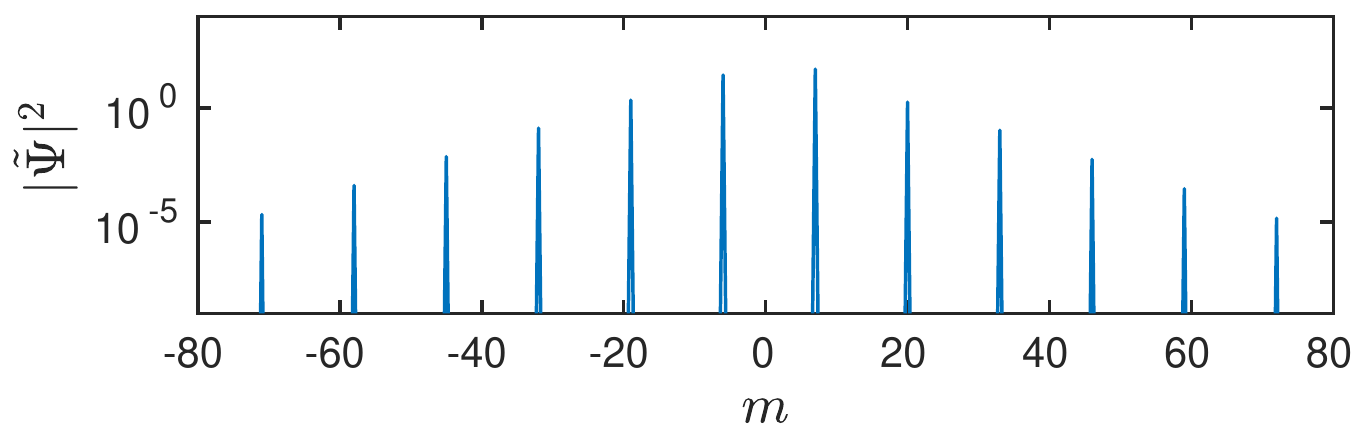}
	\caption{Same as Fig.~\ref{fig:sol1} but for $\Omega=118$, $\delta=40$ representing the sub-phase $S_{13}$. Four central peaks in $|\widetilde \Psi|^2$ are located at $ m= \{7,-6,20,-19\}$.}
	\label{fig:sol2}
\end{figure}

Examples of ground-states representing two different stripe sub-phases are shown in Figs.~\ref{fig:sol1} and \ref{fig:sol2}. The density distributions show a clear periodic structure with $k$ periods, though it is clearly different from the cosine-like shape predicted by Eq.~(\ref{eq:dens-2m}).
Similar structures were recently predicted theoretically for ultracold Fermi gases with SOAM coupling \cite{PhysRevLett.126.193401}.

In order to find the mode composition of the ground states and relate it to Eq.~(\ref{eq:nwave}) we also show in Figs.~\ref{fig:sol1} and \ref{fig:sol2} the power spectra, which can be defined for a spinor state as $|\widetilde{\Psi}(m)|^2 = |\widetilde \psi_a(m)|^2 + |\widetilde \psi_b(m)|^2$, with $\widetilde \psi_a(m)$ and $\widetilde \psi_b(m)$ being Fourier transforms of the corresponding wave functions.
By calculating such power spectra for different sub-phases we confirm that they always follow the Eq.~(\ref{eq:nwave}) with $k$ having the same meaning as in the  label of our sub-phases and $m=k/2$ for even-$k$ states, and $m=(k+1)/2$ for odd-$k$ states. One may notice from Fig.~\ref{fig:int_pd}, that with zero detuning we observe only even-$k$ states, while odd-$k$ states appear only with finite $\delta$. This is quite natural, since the superposition (\ref{eq:nwave}) can fulfill the time reversal symmetry only with even values of $k$ and with $m=k/2$.

In Fig.~\ref{fig:int_pd}(b) and (c) we also show the phase diagrams calculated within the variational approximations described in the previous section.
We see that the two-mode variational ansatz provides the phase diagram with only one stripe phase $S_{20}$ and fails to describe any of the sub-phases observed in the full calculation.
This effectively means that if we consider the variational function as a sum of two angular modes, then these two modes are always identified with $m=\pm m_0$ \cite{PhysRevA.91.063627}.
We should emphasize that parameter space of the two-mode atsatz is not restricted to $m=m_0$. Still, other values did not appear within the stripe phase during energy minimization.
 
The four-mode variational ansatz shows qualitatively some of the fine structure of the stripe phase. However, it still deviates from the correct result, especially for low-$k$ sub-phases. Importantly, the difference in the results obtained with the two-mode and the four-mode variational approximations clearly shows that the fine structure of the stripe phase is the result of admixing of higher-order angular-momentum modes in the wave function.

\begin{figure}[tbp]
	\centering
	\includegraphics[width=\linewidth]{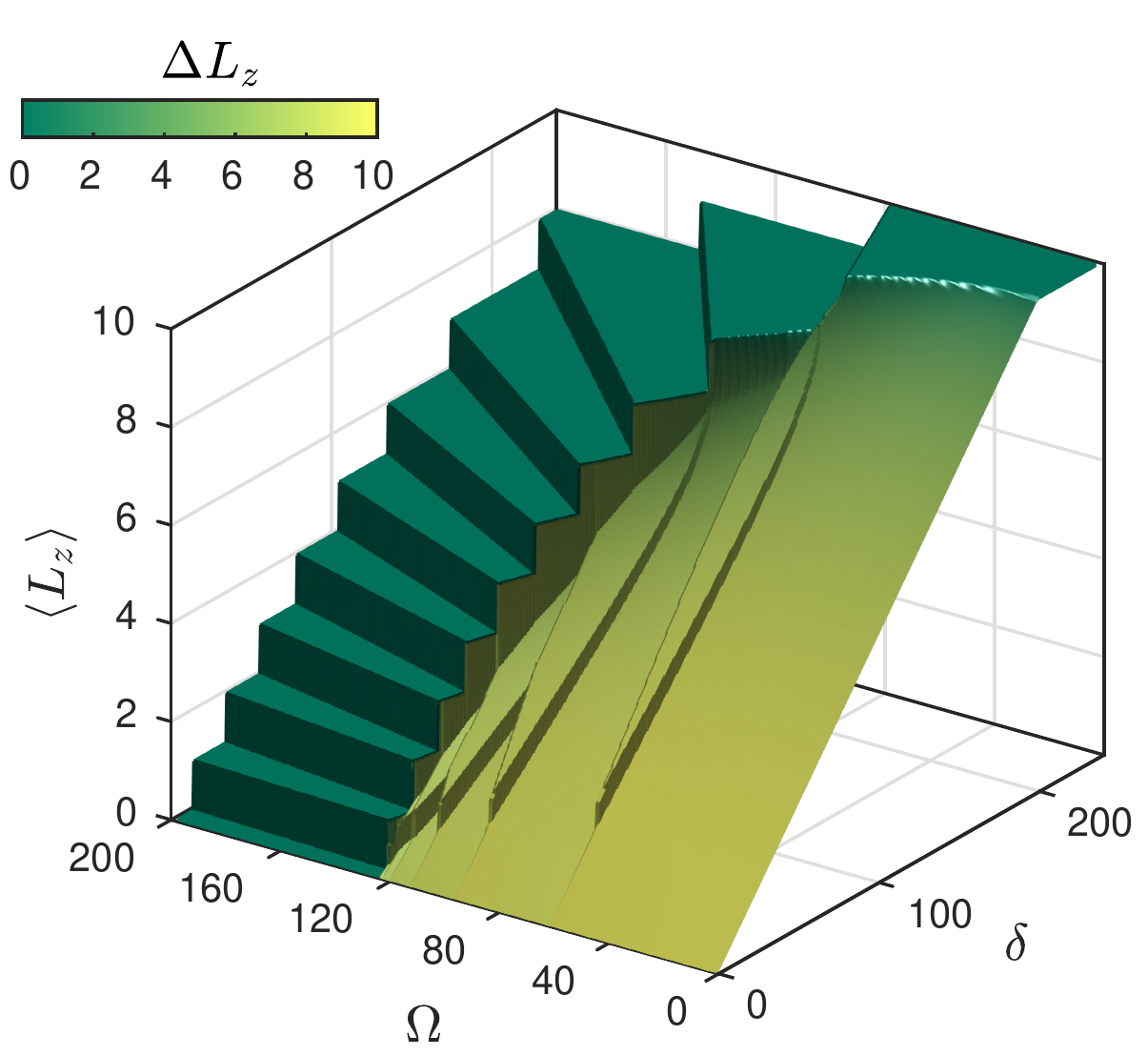}
	\caption{Expectation value of the angular momentum $\langle L_z \rangle$ and its standard deviation $\Delta L_z$ (shown with color) as functions of Raman coupling $\Omega$ and detuning $\delta$. All parameters are the same as in Fig.~\ref{fig:int_pd}}
	\label{fig:surf}
\end{figure}

All the phases and phase transitions, including internal structure of the SM and  stripe phases, can be identified from the behavior of physical observables, such as previously introduced expectation value of angular momentum $\langle L_z \rangle$ and its standard deviation $\Delta L_z$. In Fig.~\ref{fig:surf} we show these quantities calculated within the same parametric region as Fig.~\ref{fig:int_pd}.
Well defined angular momentum in the SM phase produces a characteristic
staircase-like pattern in $\langle L_z \rangle$ with zero values of $\Delta L_z$.
Within the stripe phase we observe mostly linear growth of the angular momentum with the detuning $\delta$. In the limit of small Raman coupling $\Omega$ this behavior can be deduced from Eqs.~(\ref{eq:lz-an}) and (\ref{eq:small-delta}) giving the expression $\langle L_z \rangle = m_0\delta/4G_2$, which correlates nicely with our numerical results. Interestingly, while the angular momentum is not a well-defined quantity in  the stripe phase, it still shows a discontinuous behavior, indicating first-order phase transitions between the stripe sub-phases S$_k$.

\begin{figure}[tbp]
	\centering
	\includegraphics[width=\linewidth]{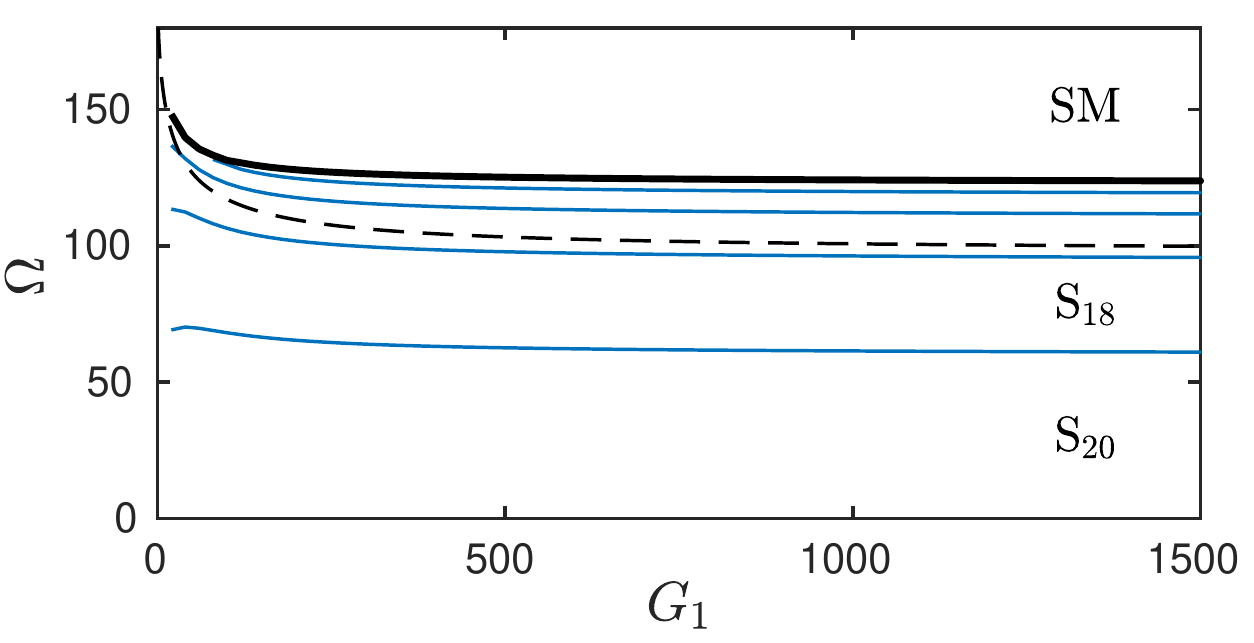}
	\includegraphics[width=\linewidth]{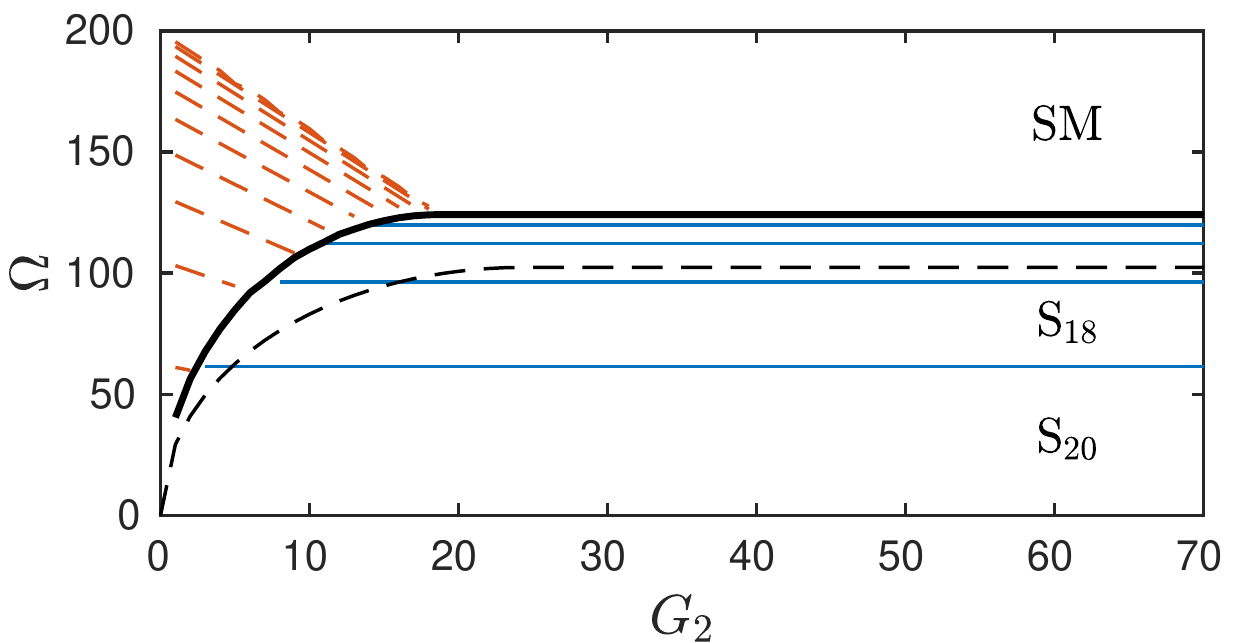}
	\caption{Top panel shows the phase diagram on the $(\Omega,G_1)$--plane with other parameters fixed as $\delta=0$ and $G_2=20$.
	Lower panel shows the phase diagram on the $(\Omega,G_2)$--plane with $\delta=0$ and $G_1=1000$. On both panels thick black line shows the boundary between the stripe phase and the single-momentum (SM) phase, blue solid lines represent boundaries of different stripe states S$_k$ (only first two are labeled) and dashed black line shows the analytical prediction of the two-mode ansatz [Eqs.~(\ref{eq:g2-cond}) -- (\ref{eq:omc-2mode2})]. Red dashed lines on the lower panel correspond to transitions between different single-momentum states. }
	\label{fig:g2dep}
\end{figure}

Next, we analyze the dependencies of phase boundaries on the nonlinear parameters $G_1$ and $G_2$. To this end Fig.~\ref{fig:g2dep} shows phase diagrams calculated within $(\Omega,G_2)$ and $(\Omega,G_1)$ planes. We restrict these phase diagrams to the region $G_1 \gg G_2$.
Comparing the obtained results with the two-mode variational atsatz we see that the phase boundary of the stripe phase can be only qualitatively reproduced by Eqs.~(\ref{eq:g2-cond}) -- (\ref{eq:omc-2mode2}). The quantitative agreement is, however, rather poor. Throughout the whole parametric range shown in the figure the stripe phase exist in considerably wider region then it is predicted by the two-mode variational approximation.

The most important observation from Fig.~\ref{fig:g2dep} is that different sub-phases S$_k$ persist within the stripe phase through a wide range of parameters $G_1$ and  $G_2$.
One may also notice that the boundaries of different stripe sub-phases S$_k$ are completely independent on $G_2$. This behavior is understandable, since with zero detuning $\delta=0$ the stripe states possess a time-reversal symmetry, implying the relation between the wave function components $|\psi_a|=|\psi_b|$, and consequently, the energy functional (\ref{eq:gpe_energy2}) becomes independent on $G_2$. Somewhat more surprising observation is that the boundaries of stripe sub-phases also become almost independent on $G_1$ in the region of large values of this interaction parameter. 
It is therefore expected that the internal fine structure of the stripe phase is quite robust with respect to the nonlinear coupling parameters and should be manifested in experimentally accessible scenarios.

\section{Application to a realistic ring system}

So far we have analyzed the internal structure of the stripe phase and have shown that this structure persists within a wide range of parameters of the system. 
In this section we test these predictions for a realistic system with experimentally realizable parameters.
As an example we consider a BEC of $^{23}$Na atoms in two hyperfine substates $|F=1,m_F=\pm1\rangle$ from the ground-state manifold. The total number of atoms in the condensate is $N=5\times10^5$ and the ring trap is defined by the potential (\ref{eq:torpot})
with $\omega_r=2\pi\times400\,\mathrm{Hz}$, $\omega_z=2\pi\times542\,\mathrm{Hz}$, $r_0=22\,\mathrm{\mu m}$, which is close to the existing experiments with sodium condensates \cite{Wang_2015}.
We now minimize a full three-dimensional Gross-Pitaevskii energy functional
\begin{multline}\label{eq:gpe_energy3d}
	E = \int d\mathbf{r} \left[  
	\begin{pmatrix}
		\psi_a^* &
		\psi_b^*
	\end{pmatrix}
	\mathcal{H}_0  
	\begin{pmatrix}
		\psi_a\\
		\psi_b
	\end{pmatrix} + \frac{g^{(\mathrm{3D})}}2 |\psi_a|^4  \right. \\ \left. + \frac{g^{(\mathrm{3D})}}2 |\psi_b|^4 + g^{(\mathrm{3D})}_{ab} |\psi_a|^2 |\psi_b|^2
	\right],
\end{multline}
with the single particle Hamiltonian from Eq.~(\ref{eq:ham0}) and the nonlinear interaction constants defined as
\[
g^{(\mathrm{3D})} = \frac{4N\pi\hbar^2}{M}a, \qquad g^{(\mathrm{3D})}_{ab} = \frac{4N\pi\hbar^2}{M}a_{ab},
\]
where $M$ is the atom mass, $a=54.54a_B$ and $a_{ab}=50.78a_B$ are s-wave scattering lengths between the same and different spin states of $^{23}$Na,  with $a_B$ being the Bohr radius \cite{PhysRevA.97.063615,PhysRevA.83.042704}. 
The Raman coupling is considered here with its radial distribution $f(r)$, which is of a Laguerre-Gaussian shape with an intensity maximum at $r_0$, so that 
\[
f(r) = A \left(r/r_0\right)^{2m_0} e^{-m_0(r/r_0)^2},
\]
where $A$ is a normalization coefficient, defined according to Eq.~(\ref{eq:rz-norm}).

The above physical parameters correspond to the reduced one-dimensional model described in Sec.~\ref{sec:model} with the energy scale $\epsilon = 2\pi/\hbar \times 0.91\,\mathrm{Hz}$ and the dimensionless nonlinear interaction coefficients $G_1=1053$ and $G_2=37.6$. 
From these values we can estimate the stripe phase region using the approximate expressions derived above using the two-mode variational model. We find, that stripe phase should be expected for $\tilde\delta<2\pi\times 136.6\,\mathrm{Hz}$ and $\tilde\Omega < 2\pi\times 93\,\mathrm{Hz}$.

\begin{figure}[tbp]
	\centering
	\includegraphics[width=\linewidth]{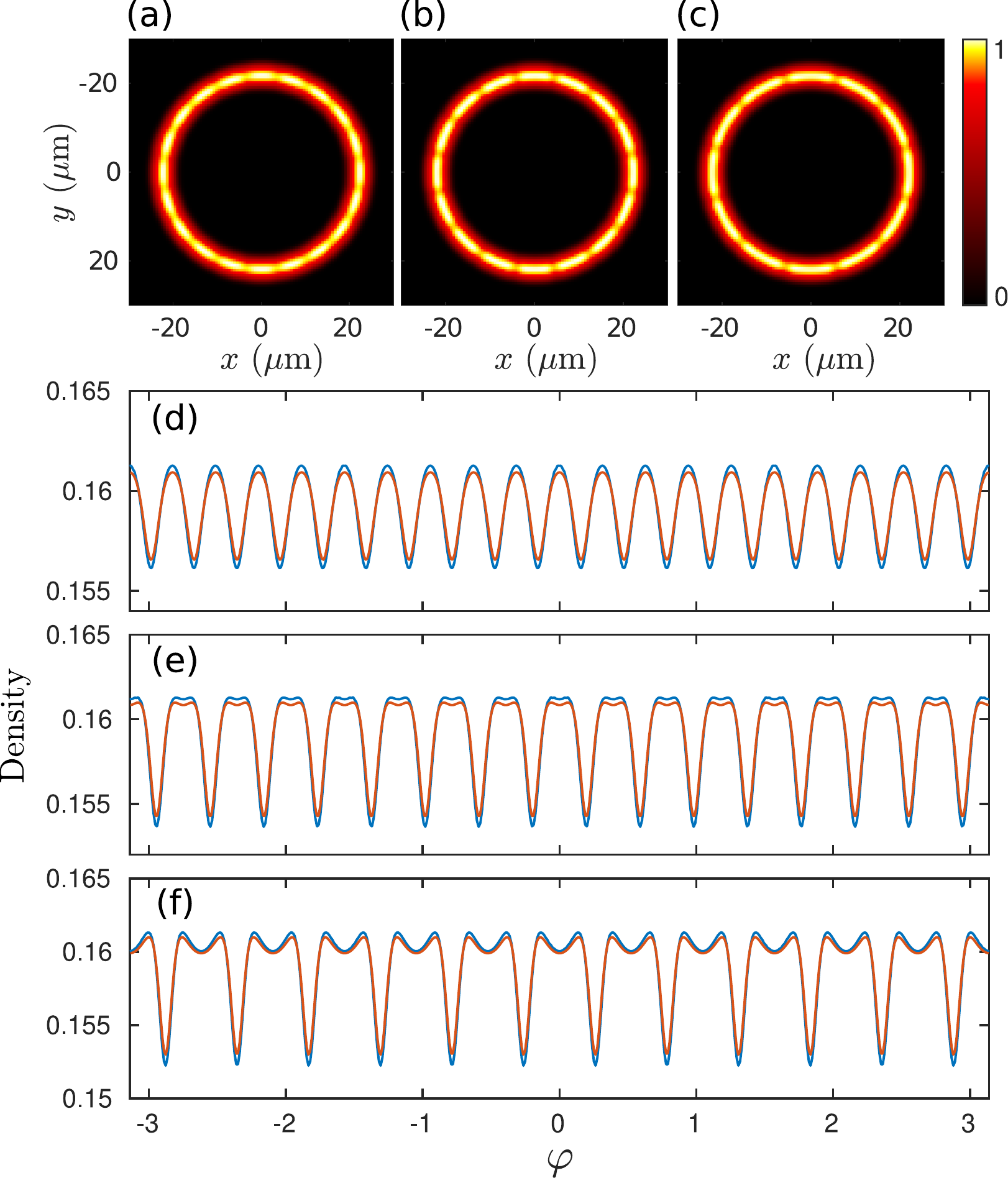}
	\caption{
Results of the ground state calculations for the example three-dimensional system
with $\tilde \delta=0$ and $\tilde \Omega = 2\pi\times 54\,\mathrm{Hz}$ [panels (a) and (d)], $\tilde \Omega = 2\pi\times 93\,\mathrm{Hz}$ [panels (b) and (e)], $\tilde \Omega = 2\pi\times 112\,\mathrm{Hz}$ [panels (c) and (f)]. Panels (a)--(c) show two-dimensional column density (in arb. units) and panels (d)--(f) show one-dimensional angular density (dimensionless). Blue lines are obtained from three-dimensional ground states by integrating out the longitudinal and radial dimensions, red lines are the results of one-dimensional model.
	}
	\label{fig:na1}
\end{figure}

The ground state of the three-dimensional system is obtained by direct numerical minimization of Eq.~(\ref{eq:gpe_energy3d}) using the gradient flow method with discrete normalization and with backward Euler pseudospectral discretization scheme \cite{doi:10.1137/140979241}. 
In Fig.~\ref{fig:na1} we show three examples of ground states obtained with $\tilde\delta=0$ and varying $\tilde\Omega$.
The values of the Raman coupling $\tilde\Omega$ are chosen such that the ground state represents different stripe sub-phases S$_{20}$, S$_{16}$ and S$_{12}$.
It is worth noticing, that the value $\tilde \Omega = 2\pi\times 112\,\mathrm{Hz}$ lies already beyond the region of stripe phase predicted by the two-mode model. This is in line with the results shown in figs.~\ref{fig:int_pd} and \ref{fig:g2dep} and confirms that the two-mode prediction is only a rough estimate of the phase boundary.
By integrating the particle density distribution over the radial and longitudinal directions we obtain a one-dimensional angular density, which show a nearly perfect agreement with the results of the one-dimensional model [see Fig.~\ref{fig:na1}(d)--(f)].
Shown examples confirm the validity of approximations used for the one-dimensional model in Sec.~\ref{sec:model} and also confirm the general claim about multiple types of stripe states in the system.

\begin{figure}[tbp]
	\centering
	\includegraphics[width=\linewidth]{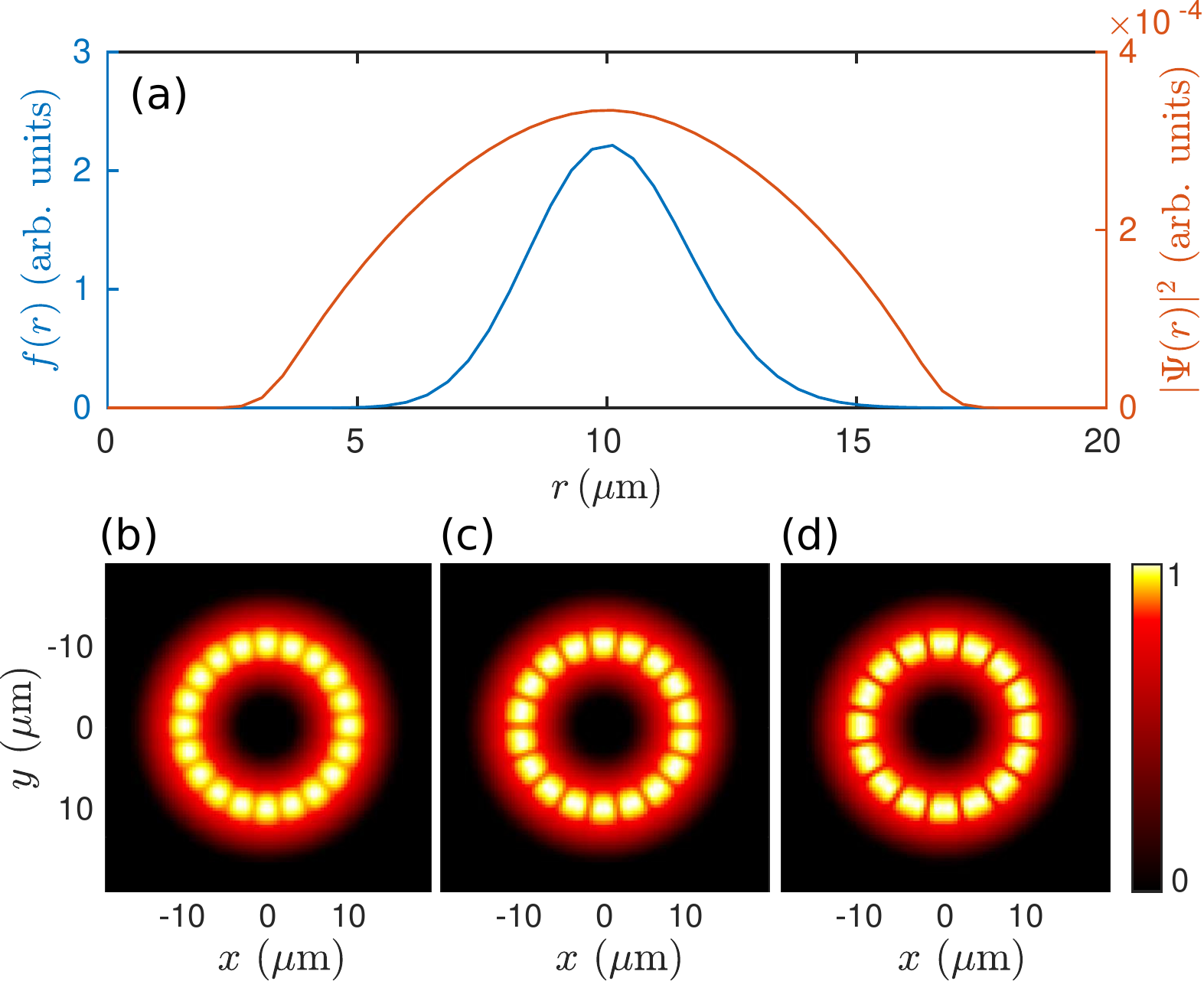}
	\caption{Results of the ground state calculations for the second example three-dimensional system. Upper panel (a) shows the radial profile of the Raman coupling (blue line and left axis) and of the condensate particle density (red line and right axis). Lower panels show calculated ground-state column densities for $\tilde \delta=0$ and $\tilde \Omega = 2\pi\times 114\,\mathrm{Hz}$ (b), $\tilde \Omega = 2\pi\times 185\,\mathrm{Hz}$ (c), $\tilde \Omega = 2\pi\times 242\,\mathrm{Hz}$ (d).
	}
	\label{fig:na2}
\end{figure}

Next, we consider a slightly different example with the ring radius reduced to $r_0=10\,\mathrm{\mu m}$, the total number of atoms increased to $N=2\times10^6$ and the other physical parameter same as above. In this case the one-dimensional model is not justified any more. Additionally, the LG profile of the Raman coupling, which we also change to have a maximum at the new $r_0$, is now considerably inhomogeneous within the condensate [see Fig.~\ref{fig:na2}(a)].
By calculating the ground states of this system with different values of $\tilde \Omega$ we can again identify stripe states with different number of density modulations [see Fig.~\ref{fig:na2}(b--d)], which shows that such features can be observed also beyond purely one-dimensional regimes.
More detailed analysis of stripe sub-phases in realistic scenarios and the effects of the trap potential will be presented in a forthcoming publication.

\section{Conclusions}

In the present work we have analyzed the structure of the ground-state phases in a toroidal two-component Bose-Einstein condensate with a spin-orbital-angular-momentum coupling between the components. In particular, a series of sub-phases is revealed inside the stripe phase of the system. These sub-phases are distinguished by the number of spatial density modulations and their characteristic power spectra. The existence of such fine structure of the stripe phase is a natural result of the wave function periodicity. 
It is worth noticing that all observed phase transitions are of the first order, which is due to quantization of the angular momentum. While this result seems natural within the SM phase, which obeys the rotational symmetry, it is somewhat surprising withing the stripe phase, where the rotational symmetry is spontaneously broken.

The obtained results are analyzed and verified through a comparison with several well established variational ans\"atze.
We observe that within a wide region of physical parameters the commonly used two-mode approximation gives only a rough description of the stripe phase of the SOAM-coupled system. Predictions of this approximation for the phase boundaries  show only limited qualitative agreement with full numerical calculations. The two-mode model also completely fails to predict the internal fine structure of the stripe phase.
These problems can be partially rectified by extending the variational model to four angular modes.
Therefore, while the two-mode approximation can provide useful analytical estimates for SOAM-coupled condensate, one should be careful with drawing physical conclusions from this approximation. 

Finally, we have demonstrated that the predicted fine structure of the stripe phase can be observed within experimentally accessible parametric regimes. To this end we have performed a full three-dimensional modeling of a realistic two-component condensate of $^{23}$Na atoms. 
We were able to observe different types of stripe states and their spatial structure agrees with predictions of the one-dimensional model.

\begin{acknowledgments}
	Y.B. acknowledges funding by the Deutsche Forschungsgemeinschaft  (DFG, German Research Foundation) -- Projektnummer 445408588.
\end{acknowledgments}

	\bibliography{refs}

\begin{thebibliography}{33}%
\makeatletter
\providecommand \@ifxundefined [1]{%
 \@ifx{#1\undefined}
}%
\providecommand \@ifnum [1]{%
 \ifnum #1\expandafter \@firstoftwo
 \else \expandafter \@secondoftwo
 \fi
}%
\providecommand \@ifx [1]{%
 \ifx #1\expandafter \@firstoftwo
 \else \expandafter \@secondoftwo
 \fi
}%
\providecommand \natexlab [1]{#1}%
\providecommand \enquote  [1]{``#1''}%
\providecommand \bibnamefont  [1]{#1}%
\providecommand \bibfnamefont [1]{#1}%
\providecommand \citenamefont [1]{#1}%
\providecommand \href@noop [0]{\@secondoftwo}%
\providecommand \href [0]{\begingroup \@sanitize@url \@href}%
\providecommand \@href[1]{\@@startlink{#1}\@@href}%
\providecommand \@@href[1]{\endgroup#1\@@endlink}%
\providecommand \@sanitize@url [0]{\catcode `\\12\catcode `\$12\catcode
  `\&12\catcode `\#12\catcode `\^12\catcode `\_12\catcode `\%12\relax}%
\providecommand \@@startlink[1]{}%
\providecommand \@@endlink[0]{}%
\providecommand \url  [0]{\begingroup\@sanitize@url \@url }%
\providecommand \@url [1]{\endgroup\@href {#1}{\urlprefix }}%
\providecommand \urlprefix  [0]{URL }%
\providecommand \Eprint [0]{\href }%
\providecommand \doibase [0]{https://doi.org/}%
\providecommand \selectlanguage [0]{\@gobble}%
\providecommand \bibinfo  [0]{\@secondoftwo}%
\providecommand \bibfield  [0]{\@secondoftwo}%
\providecommand \translation [1]{[#1]}%
\providecommand \BibitemOpen [0]{}%
\providecommand \bibitemStop [0]{}%
\providecommand \bibitemNoStop [0]{.\EOS\space}%
\providecommand \EOS [0]{\spacefactor3000\relax}%
\providecommand \BibitemShut  [1]{\csname bibitem#1\endcsname}%
\let\auto@bib@innerbib\@empty
\bibitem [{\citenamefont {Goldman}\ \emph {et~al.}(2014)\citenamefont
  {Goldman}, \citenamefont {Juzeli{\={u}}nas}, \citenamefont {Öhberg},\ and\
  \citenamefont {Spielman}}]{Goldman_2014}%
  \BibitemOpen
  \bibfield  {author} {\bibinfo {author} {\bibfnamefont {N.}~\bibnamefont
  {Goldman}}, \bibinfo {author} {\bibfnamefont {G.}~\bibnamefont
  {Juzeli{\={u}}nas}}, \bibinfo {author} {\bibfnamefont {P.}~\bibnamefont
  {Öhberg}},\ and\ \bibinfo {author} {\bibfnamefont {I.~B.}\ \bibnamefont
  {Spielman}},\ }\bibfield  {title} {\bibinfo {title} {Light-induced gauge
  fields for ultracold atoms},\ }\href
  {https://doi.org/10.1088/0034-4885/77/12/126401} {\bibfield  {journal}
  {\bibinfo  {journal} {Reports on Progress in Physics}\ }\textbf {\bibinfo
  {volume} {77}},\ \bibinfo {pages} {126401} (\bibinfo {year}
  {2014})}\BibitemShut {NoStop}%
\bibitem [{\citenamefont {Zhai}(2015)}]{Zhai2015}%
  \BibitemOpen
  \bibfield  {author} {\bibinfo {author} {\bibfnamefont {H.}~\bibnamefont
  {Zhai}},\ }\bibfield  {title} {\bibinfo {title} {Degenerate quantum gases
  with spin-orbit coupling: a review},\ }\href
  {https://doi.org/10.1088/0034-4885/78/2/026001} {\bibfield  {journal}
  {\bibinfo  {journal} {Rep. Prog. Phys.}\ }\textbf {\bibinfo {volume} {78}},\
  \bibinfo {pages} {026001} (\bibinfo {year} {2015})}\BibitemShut {NoStop}%
\bibitem [{\citenamefont {Zhang}\ and\ \citenamefont {Jo}(2019)}]{ZHANG201975}%
  \BibitemOpen
  \bibfield  {author} {\bibinfo {author} {\bibfnamefont {S.}~\bibnamefont
  {Zhang}}\ and\ \bibinfo {author} {\bibfnamefont {G.-B.}\ \bibnamefont {Jo}},\
  }\bibfield  {title} {\bibinfo {title} {Recent advances in spin-orbit coupled
  quantum gases},\ }\href
  {https://doi.org/https://doi.org/10.1016/j.jpcs.2018.04.033} {\bibfield
  {journal} {\bibinfo  {journal} {Journal of Physics and Chemistry of Solids}\
  }\textbf {\bibinfo {volume} {128}},\ \bibinfo {pages} {75 } (\bibinfo {year}
  {2019})}\BibitemShut {NoStop}%
\bibitem [{\citenamefont {Li}\ \emph {et~al.}(2017)\citenamefont {Li},
  \citenamefont {Lee}, \citenamefont {Huang}, \citenamefont {Burchesky},
  \citenamefont {Shteynas}, \citenamefont {Top}, \citenamefont {Jamison},\ and\
  \citenamefont {Ketterle}}]{Li2017}%
  \BibitemOpen
  \bibfield  {author} {\bibinfo {author} {\bibfnamefont {J.-R.}\ \bibnamefont
  {Li}}, \bibinfo {author} {\bibfnamefont {J.}~\bibnamefont {Lee}}, \bibinfo
  {author} {\bibfnamefont {W.}~\bibnamefont {Huang}}, \bibinfo {author}
  {\bibfnamefont {S.}~\bibnamefont {Burchesky}}, \bibinfo {author}
  {\bibfnamefont {B.}~\bibnamefont {Shteynas}}, \bibinfo {author}
  {\bibfnamefont {F.~{\c{C}}.}\ \bibnamefont {Top}}, \bibinfo {author}
  {\bibfnamefont {A.~O.}\ \bibnamefont {Jamison}},\ and\ \bibinfo {author}
  {\bibfnamefont {W.}~\bibnamefont {Ketterle}},\ }\bibfield  {title} {\bibinfo
  {title} {A stripe phase with supersolid properties in spin-orbit-coupled
  {Bose-Einstein} condensates},\ }\href {https://doi.org/10.1038/nature21431}
  {\bibfield  {journal} {\bibinfo  {journal} {Nature}\ }\textbf {\bibinfo
  {volume} {543}},\ \bibinfo {pages} {91} (\bibinfo {year} {2017})}\BibitemShut
  {NoStop}%
\bibitem [{\citenamefont {Boninsegni}\ and\ \citenamefont
  {Prokof'ev}(2012)}]{RevModPhys.84.759}%
  \BibitemOpen
  \bibfield  {author} {\bibinfo {author} {\bibfnamefont {M.}~\bibnamefont
  {Boninsegni}}\ and\ \bibinfo {author} {\bibfnamefont {N.~V.}\ \bibnamefont
  {Prokof'ev}},\ }\bibfield  {title} {\bibinfo {title} {{Colloquium:
  Supersolids: What and where are they?}},\ }\href
  {https://doi.org/10.1103/RevModPhys.84.759} {\bibfield  {journal} {\bibinfo
  {journal} {Rev. Mod. Phys.}\ }\textbf {\bibinfo {volume} {84}},\ \bibinfo
  {pages} {759} (\bibinfo {year} {2012})}\BibitemShut {NoStop}%
\bibitem [{\citenamefont {Lin}\ \emph {et~al.}(2011)\citenamefont {Lin},
  \citenamefont {Jim{\'e}nez-Garc{\'i}a},\ and\ \citenamefont
  {Spielman}}]{Lin2011}%
  \BibitemOpen
  \bibfield  {author} {\bibinfo {author} {\bibfnamefont {Y.-J.}\ \bibnamefont
  {Lin}}, \bibinfo {author} {\bibfnamefont {K.}~\bibnamefont
  {Jim{\'e}nez-Garc{\'i}a}},\ and\ \bibinfo {author} {\bibfnamefont {I.~B.}\
  \bibnamefont {Spielman}},\ }\bibfield  {title} {\bibinfo {title}
  {Spin-orbit-coupled {Bose-Einstein} condensates},\ }\href
  {https://doi.org/10.1038/nature09887} {\bibfield  {journal} {\bibinfo
  {journal} {Nature}\ }\textbf {\bibinfo {volume} {471}},\ \bibinfo {pages}
  {83} (\bibinfo {year} {2011})}\BibitemShut {NoStop}%
\bibitem [{\citenamefont {Wang}\ \emph {et~al.}(2012)\citenamefont {Wang},
  \citenamefont {Yu}, \citenamefont {Fu}, \citenamefont {Miao}, \citenamefont
  {Huang}, \citenamefont {Chai}, \citenamefont {Zhai},\ and\ \citenamefont
  {Zhang}}]{PhysRevLett.109.095301}%
  \BibitemOpen
  \bibfield  {author} {\bibinfo {author} {\bibfnamefont {P.}~\bibnamefont
  {Wang}}, \bibinfo {author} {\bibfnamefont {Z.-Q.}\ \bibnamefont {Yu}},
  \bibinfo {author} {\bibfnamefont {Z.}~\bibnamefont {Fu}}, \bibinfo {author}
  {\bibfnamefont {J.}~\bibnamefont {Miao}}, \bibinfo {author} {\bibfnamefont
  {L.}~\bibnamefont {Huang}}, \bibinfo {author} {\bibfnamefont
  {S.}~\bibnamefont {Chai}}, \bibinfo {author} {\bibfnamefont {H.}~\bibnamefont
  {Zhai}},\ and\ \bibinfo {author} {\bibfnamefont {J.}~\bibnamefont {Zhang}},\
  }\bibfield  {title} {\bibinfo {title} {Spin-orbit coupled degenerate {Fermi}
  gases},\ }\href {https://doi.org/10.1103/PhysRevLett.109.095301} {\bibfield
  {journal} {\bibinfo  {journal} {Phys. Rev. Lett.}\ }\textbf {\bibinfo
  {volume} {109}},\ \bibinfo {pages} {095301} (\bibinfo {year}
  {2012})}\BibitemShut {NoStop}%
\bibitem [{\citenamefont {Luo}\ \emph {et~al.}(2016)\citenamefont {Luo},
  \citenamefont {Wu}, \citenamefont {Chen}, \citenamefont {Guan}, \citenamefont
  {Gao}, \citenamefont {Xu}, \citenamefont {You},\ and\ \citenamefont
  {Wang}}]{luo2016tunable}%
  \BibitemOpen
  \bibfield  {author} {\bibinfo {author} {\bibfnamefont {X.}~\bibnamefont
  {Luo}}, \bibinfo {author} {\bibfnamefont {L.}~\bibnamefont {Wu}}, \bibinfo
  {author} {\bibfnamefont {J.}~\bibnamefont {Chen}}, \bibinfo {author}
  {\bibfnamefont {Q.}~\bibnamefont {Guan}}, \bibinfo {author} {\bibfnamefont
  {K.}~\bibnamefont {Gao}}, \bibinfo {author} {\bibfnamefont {Z.-F.}\
  \bibnamefont {Xu}}, \bibinfo {author} {\bibfnamefont {L.}~\bibnamefont
  {You}},\ and\ \bibinfo {author} {\bibfnamefont {R.}~\bibnamefont {Wang}},\
  }\bibfield  {title} {\bibinfo {title} {Tunable atomic spin-orbit coupling
  synthesized with a modulating gradient magnetic field},\ }\href
  {https://doi.org/10.1038/srep18983} {\bibfield  {journal} {\bibinfo
  {journal} {Scientific reports}\ }\textbf {\bibinfo {volume} {6}},\ \bibinfo
  {pages} {1} (\bibinfo {year} {2016})}\BibitemShut {NoStop}%
\bibitem [{\citenamefont {Li}\ \emph {et~al.}(2016)\citenamefont {Li},
  \citenamefont {Huang}, \citenamefont {Shteynas}, \citenamefont {Burchesky},
  \citenamefont {Top}, \citenamefont {Su}, \citenamefont {Lee}, \citenamefont
  {Jamison},\ and\ \citenamefont {Ketterle}}]{PhysRevLett.117.185301}%
  \BibitemOpen
  \bibfield  {author} {\bibinfo {author} {\bibfnamefont {J.-R.}\ \bibnamefont
  {Li}}, \bibinfo {author} {\bibfnamefont {W.}~\bibnamefont {Huang}}, \bibinfo
  {author} {\bibfnamefont {B.}~\bibnamefont {Shteynas}}, \bibinfo {author}
  {\bibfnamefont {S.}~\bibnamefont {Burchesky}}, \bibinfo {author}
  {\bibfnamefont {F.~{\c{C}}.}\ \bibnamefont {Top}}, \bibinfo {author}
  {\bibfnamefont {E.}~\bibnamefont {Su}}, \bibinfo {author} {\bibfnamefont
  {J.}~\bibnamefont {Lee}}, \bibinfo {author} {\bibfnamefont {A.~O.}\
  \bibnamefont {Jamison}},\ and\ \bibinfo {author} {\bibfnamefont
  {W.}~\bibnamefont {Ketterle}},\ }\bibfield  {title} {\bibinfo {title}
  {Spin-orbit coupling and spin textures in optical superlattices},\ }\href
  {https://doi.org/10.1103/PhysRevLett.117.185301} {\bibfield  {journal}
  {\bibinfo  {journal} {Phys. Rev. Lett.}\ }\textbf {\bibinfo {volume} {117}},\
  \bibinfo {pages} {185301} (\bibinfo {year} {2016})}\BibitemShut {NoStop}%
\bibitem [{\citenamefont {Wu}\ \emph {et~al.}(2016)\citenamefont {Wu},
  \citenamefont {Zhang}, \citenamefont {Sun}, \citenamefont {Xu}, \citenamefont
  {Wang}, \citenamefont {Ji}, \citenamefont {Deng}, \citenamefont {Chen},
  \citenamefont {Liu},\ and\ \citenamefont {Pan}}]{Wu2016}%
  \BibitemOpen
  \bibfield  {author} {\bibinfo {author} {\bibfnamefont {Z.}~\bibnamefont
  {Wu}}, \bibinfo {author} {\bibfnamefont {L.}~\bibnamefont {Zhang}}, \bibinfo
  {author} {\bibfnamefont {W.}~\bibnamefont {Sun}}, \bibinfo {author}
  {\bibfnamefont {X.-T.}\ \bibnamefont {Xu}}, \bibinfo {author} {\bibfnamefont
  {B.-Z.}\ \bibnamefont {Wang}}, \bibinfo {author} {\bibfnamefont {S.-C.}\
  \bibnamefont {Ji}}, \bibinfo {author} {\bibfnamefont {Y.}~\bibnamefont
  {Deng}}, \bibinfo {author} {\bibfnamefont {S.}~\bibnamefont {Chen}}, \bibinfo
  {author} {\bibfnamefont {X.-J.}\ \bibnamefont {Liu}},\ and\ \bibinfo {author}
  {\bibfnamefont {J.-W.}\ \bibnamefont {Pan}},\ }\bibfield  {title} {\bibinfo
  {title} {Realization of two-dimensional spin-orbit coupling for
  {Bose-Einstein} condensates},\ }\href
  {https://doi.org/10.1126/science.aaf6689} {\bibfield  {journal} {\bibinfo
  {journal} {Science}\ }\textbf {\bibinfo {volume} {354}},\ \bibinfo {pages}
  {83} (\bibinfo {year} {2016})}\BibitemShut {NoStop}%
\bibitem [{\citenamefont {Kartashov}\ and\ \citenamefont
  {Konotop}(2017)}]{PhysRevLett.118.190401}%
  \BibitemOpen
  \bibfield  {author} {\bibinfo {author} {\bibfnamefont {Y.~V.}\ \bibnamefont
  {Kartashov}}\ and\ \bibinfo {author} {\bibfnamefont {V.~V.}\ \bibnamefont
  {Konotop}},\ }\bibfield  {title} {\bibinfo {title} {{Solitons in
  Bose-Einstein Condensates with Helicoidal Spin-Orbit Coupling}},\ }\href
  {https://doi.org/10.1103/PhysRevLett.118.190401} {\bibfield  {journal}
  {\bibinfo  {journal} {Phys. Rev. Lett.}\ }\textbf {\bibinfo {volume} {118}},\
  \bibinfo {pages} {190401} (\bibinfo {year} {2017})}\BibitemShut {NoStop}%
\bibitem [{\citenamefont {Chen}\ \emph {et~al.}(2018)\citenamefont {Chen},
  \citenamefont {Lin}, \citenamefont {Chen}, \citenamefont {Chiu},
  \citenamefont {Wang}, \citenamefont {Chen}, \citenamefont {Huang},
  \citenamefont {Yip}, \citenamefont {Kawaguchi},\ and\ \citenamefont
  {Lin}}]{PhysRevLett.121.113204}%
  \BibitemOpen
  \bibfield  {author} {\bibinfo {author} {\bibfnamefont {H.-R.}\ \bibnamefont
  {Chen}}, \bibinfo {author} {\bibfnamefont {K.-Y.}\ \bibnamefont {Lin}},
  \bibinfo {author} {\bibfnamefont {P.-K.}\ \bibnamefont {Chen}}, \bibinfo
  {author} {\bibfnamefont {N.-C.}\ \bibnamefont {Chiu}}, \bibinfo {author}
  {\bibfnamefont {J.-B.}\ \bibnamefont {Wang}}, \bibinfo {author}
  {\bibfnamefont {C.-A.}\ \bibnamefont {Chen}}, \bibinfo {author}
  {\bibfnamefont {P.-P.}\ \bibnamefont {Huang}}, \bibinfo {author}
  {\bibfnamefont {S.-K.}\ \bibnamefont {Yip}}, \bibinfo {author} {\bibfnamefont
  {Y.}~\bibnamefont {Kawaguchi}},\ and\ \bibinfo {author} {\bibfnamefont
  {Y.-J.}\ \bibnamefont {Lin}},\ }\bibfield  {title} {\bibinfo {title}
  {Spin--orbital-angular-momentum coupled {Bose-Einstein} condensates},\ }\href
  {https://doi.org/10.1103/PhysRevLett.121.113204} {\bibfield  {journal}
  {\bibinfo  {journal} {Phys. Rev. Lett.}\ }\textbf {\bibinfo {volume} {121}},\
  \bibinfo {pages} {113204} (\bibinfo {year} {2018})}\BibitemShut {NoStop}%
\bibitem [{\citenamefont {Zhang}\ \emph {et~al.}(2019)\citenamefont {Zhang},
  \citenamefont {Gao}, \citenamefont {Zou}, \citenamefont {Kong}, \citenamefont
  {Li}, \citenamefont {Shen}, \citenamefont {Chen}, \citenamefont {Peng},
  \citenamefont {Zhan}, \citenamefont {Pu},\ and\ \citenamefont
  {Jiang}}]{PhysRevLett.122.110402}%
  \BibitemOpen
  \bibfield  {author} {\bibinfo {author} {\bibfnamefont {D.}~\bibnamefont
  {Zhang}}, \bibinfo {author} {\bibfnamefont {T.}~\bibnamefont {Gao}}, \bibinfo
  {author} {\bibfnamefont {P.}~\bibnamefont {Zou}}, \bibinfo {author}
  {\bibfnamefont {L.}~\bibnamefont {Kong}}, \bibinfo {author} {\bibfnamefont
  {R.}~\bibnamefont {Li}}, \bibinfo {author} {\bibfnamefont {X.}~\bibnamefont
  {Shen}}, \bibinfo {author} {\bibfnamefont {X.-L.}\ \bibnamefont {Chen}},
  \bibinfo {author} {\bibfnamefont {S.-G.}\ \bibnamefont {Peng}}, \bibinfo
  {author} {\bibfnamefont {M.}~\bibnamefont {Zhan}}, \bibinfo {author}
  {\bibfnamefont {H.}~\bibnamefont {Pu}},\ and\ \bibinfo {author}
  {\bibfnamefont {K.}~\bibnamefont {Jiang}},\ }\bibfield  {title} {\bibinfo
  {title} {Ground-state phase diagram of a spin-orbital-angular-momentum
  coupled {Bose-Einstein} condensate},\ }\href
  {https://doi.org/10.1103/PhysRevLett.122.110402} {\bibfield  {journal}
  {\bibinfo  {journal} {Phys. Rev. Lett.}\ }\textbf {\bibinfo {volume} {122}},\
  \bibinfo {pages} {110402} (\bibinfo {year} {2019})}\BibitemShut {NoStop}%
\bibitem [{\citenamefont {Chen}\ \emph {et~al.}(2016)\citenamefont {Chen},
  \citenamefont {Pu},\ and\ \citenamefont {Zhang}}]{PhysRevA.93.013629}%
  \BibitemOpen
  \bibfield  {author} {\bibinfo {author} {\bibfnamefont {L.}~\bibnamefont
  {Chen}}, \bibinfo {author} {\bibfnamefont {H.}~\bibnamefont {Pu}},\ and\
  \bibinfo {author} {\bibfnamefont {Y.}~\bibnamefont {Zhang}},\ }\bibfield
  {title} {\bibinfo {title} {Spin-orbit angular momentum coupling in a spin-1
  {Bose-Einstein} condensate},\ }\href
  {https://doi.org/10.1103/PhysRevA.93.013629} {\bibfield  {journal} {\bibinfo
  {journal} {Phys. Rev. A}\ }\textbf {\bibinfo {volume} {93}},\ \bibinfo
  {pages} {013629} (\bibinfo {year} {2016})}\BibitemShut {NoStop}%
\bibitem [{\citenamefont {{Vasi\'{c}}}\ and\ \citenamefont
  {{Bala\v{z}}}(2016)}]{PhysRevA.94.033627}%
  \BibitemOpen
  \bibfield  {author} {\bibinfo {author} {\bibfnamefont {I.}~\bibnamefont
  {{Vasi\'{c}}}}\ and\ \bibinfo {author} {\bibfnamefont {A.}~\bibnamefont
  {{Bala\v{z}}}},\ }\bibfield  {title} {\bibinfo {title} {{Excitation spectra
  of a Bose-Einstein condensate with an angular spin-orbit coupling}},\ }\href
  {https://doi.org/10.1103/PhysRevA.94.033627} {\bibfield  {journal} {\bibinfo
  {journal} {Phys. Rev. A}\ }\textbf {\bibinfo {volume} {94}},\ \bibinfo
  {pages} {033627} (\bibinfo {year} {2016})}\BibitemShut {NoStop}%
\bibitem [{\citenamefont {Sun}\ \emph {et~al.}(2015)\citenamefont {Sun},
  \citenamefont {Qu},\ and\ \citenamefont {Zhang}}]{PhysRevA.91.063627}%
  \BibitemOpen
  \bibfield  {author} {\bibinfo {author} {\bibfnamefont {K.}~\bibnamefont
  {Sun}}, \bibinfo {author} {\bibfnamefont {C.}~\bibnamefont {Qu}},\ and\
  \bibinfo {author} {\bibfnamefont {C.}~\bibnamefont {Zhang}},\ }\bibfield
  {title} {\bibinfo {title} {Spin--orbital-angular-momentum coupling in
  {Bose-Einstein} condensates},\ }\href
  {https://doi.org/10.1103/PhysRevA.91.063627} {\bibfield  {journal} {\bibinfo
  {journal} {Phys. Rev. A}\ }\textbf {\bibinfo {volume} {91}},\ \bibinfo
  {pages} {063627} (\bibinfo {year} {2015})}\BibitemShut {NoStop}%
\bibitem [{\citenamefont {Chen}\ \emph {et~al.}(2020)\citenamefont {Chen},
  \citenamefont {Peng}, \citenamefont {Zou}, \citenamefont {Liu},\ and\
  \citenamefont {Hu}}]{PhysRevResearch.2.033152}%
  \BibitemOpen
  \bibfield  {author} {\bibinfo {author} {\bibfnamefont {X.-L.}\ \bibnamefont
  {Chen}}, \bibinfo {author} {\bibfnamefont {S.-G.}\ \bibnamefont {Peng}},
  \bibinfo {author} {\bibfnamefont {P.}~\bibnamefont {Zou}}, \bibinfo {author}
  {\bibfnamefont {X.-J.}\ \bibnamefont {Liu}},\ and\ \bibinfo {author}
  {\bibfnamefont {H.}~\bibnamefont {Hu}},\ }\bibfield  {title} {\bibinfo
  {title} {Angular stripe phase in spin-orbital-angular-momentum coupled {Bose}
  condensates},\ }\href {https://doi.org/10.1103/PhysRevResearch.2.033152}
  {\bibfield  {journal} {\bibinfo  {journal} {Phys. Rev. Research}\ }\textbf
  {\bibinfo {volume} {2}},\ \bibinfo {pages} {033152} (\bibinfo {year}
  {2020})}\BibitemShut {NoStop}%
\bibitem [{\citenamefont {Chiu}\ \emph {et~al.}(2020)\citenamefont {Chiu},
  \citenamefont {Kawaguchi}, \citenamefont {Yip},\ and\ \citenamefont
  {Lin}}]{Chiu_2020}%
  \BibitemOpen
  \bibfield  {author} {\bibinfo {author} {\bibfnamefont {N.-C.}\ \bibnamefont
  {Chiu}}, \bibinfo {author} {\bibfnamefont {Y.}~\bibnamefont {Kawaguchi}},
  \bibinfo {author} {\bibfnamefont {S.-K.}\ \bibnamefont {Yip}},\ and\ \bibinfo
  {author} {\bibfnamefont {Y.-J.}\ \bibnamefont {Lin}},\ }\bibfield  {title}
  {\bibinfo {title} {Visible stripe phases in
  spin{\textendash}orbital-angular-momentum coupled {Bose{\textendash}Einstein}
  condensates},\ }\href {https://doi.org/10.1088/1367-2630/abac3c} {\bibfield
  {journal} {\bibinfo  {journal} {New Journal of Physics}\ }\textbf {\bibinfo
  {volume} {22}},\ \bibinfo {pages} {093017} (\bibinfo {year}
  {2020})}\BibitemShut {NoStop}%
\bibitem [{\citenamefont {DeMarco}\ and\ \citenamefont
  {Pu}(2015)}]{PhysRevA.91.033630}%
  \BibitemOpen
  \bibfield  {author} {\bibinfo {author} {\bibfnamefont {M.}~\bibnamefont
  {DeMarco}}\ and\ \bibinfo {author} {\bibfnamefont {H.}~\bibnamefont {Pu}},\
  }\bibfield  {title} {\bibinfo {title} {Angular spin-orbit coupling in cold
  atoms},\ }\href {https://doi.org/10.1103/PhysRevA.91.033630} {\bibfield
  {journal} {\bibinfo  {journal} {Phys. Rev. A}\ }\textbf {\bibinfo {volume}
  {91}},\ \bibinfo {pages} {033630} (\bibinfo {year} {2015})}\BibitemShut
  {NoStop}%
\bibitem [{\citenamefont {Hu}\ \emph {et~al.}(2015)\citenamefont {Hu},
  \citenamefont {Miniatura},\ and\ \citenamefont
  {Gr\'emaud}}]{PhysRevA.92.033615}%
  \BibitemOpen
  \bibfield  {author} {\bibinfo {author} {\bibfnamefont {Y.-X.}\ \bibnamefont
  {Hu}}, \bibinfo {author} {\bibfnamefont {C.}~\bibnamefont {Miniatura}},\ and\
  \bibinfo {author} {\bibfnamefont {B.}~\bibnamefont {Gr\'emaud}},\ }\bibfield
  {title} {\bibinfo {title} {Half-skyrmion and vortex-antivortex pairs in
  spinor condensates},\ }\href {https://doi.org/10.1103/PhysRevA.92.033615}
  {\bibfield  {journal} {\bibinfo  {journal} {Phys. Rev. A}\ }\textbf {\bibinfo
  {volume} {92}},\ \bibinfo {pages} {033615} (\bibinfo {year}
  {2015})}\BibitemShut {NoStop}%
\bibitem [{\citenamefont {Duan}\ \emph {et~al.}(2020)\citenamefont {Duan},
  \citenamefont {Bidasyuk},\ and\ \citenamefont
  {Surzhykov}}]{PhysRevA.102.063328}%
  \BibitemOpen
  \bibfield  {author} {\bibinfo {author} {\bibfnamefont {Y.}~\bibnamefont
  {Duan}}, \bibinfo {author} {\bibfnamefont {Y.~M.}\ \bibnamefont {Bidasyuk}},\
  and\ \bibinfo {author} {\bibfnamefont {A.}~\bibnamefont {Surzhykov}},\
  }\bibfield  {title} {\bibinfo {title} {{Symmetry breaking and phase
  transitions in Bose-Einstein condensates with spin--orbital-angular-momentum
  coupling}},\ }\href {https://doi.org/10.1103/PhysRevA.102.063328} {\bibfield
  {journal} {\bibinfo  {journal} {Phys. Rev. A}\ }\textbf {\bibinfo {volume}
  {102}},\ \bibinfo {pages} {063328} (\bibinfo {year} {2020})}\BibitemShut
  {NoStop}%
\bibitem [{\citenamefont {Campbell}\ \emph {et~al.}(2012)\citenamefont
  {Campbell}, \citenamefont {Hage}, \citenamefont {Buchler},\ and\
  \citenamefont {Lam}}]{Campbell:12}%
  \BibitemOpen
  \bibfield  {author} {\bibinfo {author} {\bibfnamefont {G.}~\bibnamefont
  {Campbell}}, \bibinfo {author} {\bibfnamefont {B.}~\bibnamefont {Hage}},
  \bibinfo {author} {\bibfnamefont {B.}~\bibnamefont {Buchler}},\ and\ \bibinfo
  {author} {\bibfnamefont {P.~K.}\ \bibnamefont {Lam}},\ }\bibfield  {title}
  {\bibinfo {title} {Generation of high-order optical vortices using directly
  machined spiral phase mirrors},\ }\href
  {https://doi.org/10.1364/AO.51.000873} {\bibfield  {journal} {\bibinfo
  {journal} {Appl. Opt.}\ }\textbf {\bibinfo {volume} {51}},\ \bibinfo {pages}
  {873} (\bibinfo {year} {2012})}\BibitemShut {NoStop}%
\bibitem [{\citenamefont {Fickler}\ \emph {et~al.}(2016)\citenamefont
  {Fickler}, \citenamefont {Campbell}, \citenamefont {Buchler}, \citenamefont
  {Lam},\ and\ \citenamefont {Zeilinger}}]{Fickler13642}%
  \BibitemOpen
  \bibfield  {author} {\bibinfo {author} {\bibfnamefont {R.}~\bibnamefont
  {Fickler}}, \bibinfo {author} {\bibfnamefont {G.}~\bibnamefont {Campbell}},
  \bibinfo {author} {\bibfnamefont {B.}~\bibnamefont {Buchler}}, \bibinfo
  {author} {\bibfnamefont {P.~K.}\ \bibnamefont {Lam}},\ and\ \bibinfo {author}
  {\bibfnamefont {A.}~\bibnamefont {Zeilinger}},\ }\bibfield  {title} {\bibinfo
  {title} {Quantum entanglement of angular momentum states with quantum numbers
  up to 10,010},\ }\href {https://doi.org/10.1073/pnas.1616889113} {\bibfield
  {journal} {\bibinfo  {journal} {Proceedings of the National Academy of
  Sciences}\ }\textbf {\bibinfo {volume} {113}},\ \bibinfo {pages} {13642}
  (\bibinfo {year} {2016})}\BibitemShut {NoStop}%
\bibitem [{\citenamefont {Hou}\ \emph {et~al.}(2017)\citenamefont {Hou},
  \citenamefont {Luo}, \citenamefont {Sun},\ and\ \citenamefont
  {Zhang}}]{PhysRevA.96.011603}%
  \BibitemOpen
  \bibfield  {author} {\bibinfo {author} {\bibfnamefont {J.}~\bibnamefont
  {Hou}}, \bibinfo {author} {\bibfnamefont {X.-W.}\ \bibnamefont {Luo}},
  \bibinfo {author} {\bibfnamefont {K.}~\bibnamefont {Sun}},\ and\ \bibinfo
  {author} {\bibfnamefont {C.}~\bibnamefont {Zhang}},\ }\bibfield  {title}
  {\bibinfo {title} {{Adiabatically tuning quantized supercurrents in an
  annular Bose-Einstein condensate}},\ }\href
  {https://doi.org/10.1103/PhysRevA.96.011603} {\bibfield  {journal} {\bibinfo
  {journal} {Phys. Rev. A}\ }\textbf {\bibinfo {volume} {96}},\ \bibinfo
  {pages} {011603(R)} (\bibinfo {year} {2017})}\BibitemShut {NoStop}%
\bibitem [{\citenamefont {Baharian}\ and\ \citenamefont
  {Baym}(2013)}]{PhysRevA.87.013619}%
  \BibitemOpen
  \bibfield  {author} {\bibinfo {author} {\bibfnamefont {S.}~\bibnamefont
  {Baharian}}\ and\ \bibinfo {author} {\bibfnamefont {G.}~\bibnamefont
  {Baym}},\ }\bibfield  {title} {\bibinfo {title} {{Bose-Einstein condensates
  in toroidal traps: Instabilities, swallow-tail loops, and self-trapping}},\
  }\href {https://doi.org/10.1103/PhysRevA.87.013619} {\bibfield  {journal}
  {\bibinfo  {journal} {Phys. Rev. A}\ }\textbf {\bibinfo {volume} {87}},\
  \bibinfo {pages} {013619} (\bibinfo {year} {2013})}\BibitemShut {NoStop}%
\bibitem [{\citenamefont {Li}\ \emph {et~al.}(2012)\citenamefont {Li},
  \citenamefont {Pitaevskii},\ and\ \citenamefont
  {Stringari}}]{PhysRevLett.108.225301}%
  \BibitemOpen
  \bibfield  {author} {\bibinfo {author} {\bibfnamefont {Y.}~\bibnamefont
  {Li}}, \bibinfo {author} {\bibfnamefont {L.~P.}\ \bibnamefont {Pitaevskii}},\
  and\ \bibinfo {author} {\bibfnamefont {S.}~\bibnamefont {Stringari}},\
  }\bibfield  {title} {\bibinfo {title} {Quantum tricriticality and phase
  transitions in spin-orbit coupled {Bose-Einstein} condensates},\ }\href
  {https://doi.org/10.1103/PhysRevLett.108.225301} {\bibfield  {journal}
  {\bibinfo  {journal} {Phys. Rev. Lett.}\ }\textbf {\bibinfo {volume} {108}},\
  \bibinfo {pages} {225301} (\bibinfo {year} {2012})}\BibitemShut {NoStop}%
\bibitem [{\citenamefont {Qu}\ \emph {et~al.}(2016)\citenamefont {Qu},
  \citenamefont {Pitaevskii},\ and\ \citenamefont
  {Stringari}}]{PhysRevLett.116.160402}%
  \BibitemOpen
  \bibfield  {author} {\bibinfo {author} {\bibfnamefont {C.}~\bibnamefont
  {Qu}}, \bibinfo {author} {\bibfnamefont {L.~P.}\ \bibnamefont {Pitaevskii}},\
  and\ \bibinfo {author} {\bibfnamefont {S.}~\bibnamefont {Stringari}},\
  }\bibfield  {title} {\bibinfo {title} {Magnetic solitons in a binary
  {Bose-Einstein} condensate},\ }\href
  {https://doi.org/10.1103/PhysRevLett.116.160402} {\bibfield  {journal}
  {\bibinfo  {journal} {Phys. Rev. Lett.}\ }\textbf {\bibinfo {volume} {116}},\
  \bibinfo {pages} {160402} (\bibinfo {year} {2016})}\BibitemShut {NoStop}%
\bibitem [{\citenamefont {Li}\ \emph {et~al.}(2013)\citenamefont {Li},
  \citenamefont {Martone}, \citenamefont {Pitaevskii},\ and\ \citenamefont
  {Stringari}}]{PhysRevLett.110.235302}%
  \BibitemOpen
  \bibfield  {author} {\bibinfo {author} {\bibfnamefont {Y.}~\bibnamefont
  {Li}}, \bibinfo {author} {\bibfnamefont {G.~I.}\ \bibnamefont {Martone}},
  \bibinfo {author} {\bibfnamefont {L.~P.}\ \bibnamefont {Pitaevskii}},\ and\
  \bibinfo {author} {\bibfnamefont {S.}~\bibnamefont {Stringari}},\ }\bibfield
  {title} {\bibinfo {title} {Superstripes and the excitation spectrum of a
  spin-orbit-coupled {Bose-Einstein} condensate},\ }\href
  {https://doi.org/10.1103/PhysRevLett.110.235302} {\bibfield  {journal}
  {\bibinfo  {journal} {Phys. Rev. Lett.}\ }\textbf {\bibinfo {volume} {110}},\
  \bibinfo {pages} {235302} (\bibinfo {year} {2013})}\BibitemShut {NoStop}%
\bibitem [{\citenamefont {Wang}\ \emph {et~al.}(2021)\citenamefont {Wang},
  \citenamefont {Ji}, \citenamefont {Sun},\ and\ \citenamefont
  {Li}}]{PhysRevLett.126.193401}%
  \BibitemOpen
  \bibfield  {author} {\bibinfo {author} {\bibfnamefont {L.-L.}\ \bibnamefont
  {Wang}}, \bibinfo {author} {\bibfnamefont {A.-C.}\ \bibnamefont {Ji}},
  \bibinfo {author} {\bibfnamefont {Q.}~\bibnamefont {Sun}},\ and\ \bibinfo
  {author} {\bibfnamefont {J.}~\bibnamefont {Li}},\ }\bibfield  {title}
  {\bibinfo {title} {Exotic vortex states with discrete rotational symmetry in
  atomic {Fermi} gases with spin-orbital--angular-momentum coupling},\ }\href
  {https://doi.org/10.1103/PhysRevLett.126.193401} {\bibfield  {journal}
  {\bibinfo  {journal} {Phys. Rev. Lett.}\ }\textbf {\bibinfo {volume} {126}},\
  \bibinfo {pages} {193401} (\bibinfo {year} {2021})}\BibitemShut {NoStop}%
\bibitem [{\citenamefont {Wang}\ \emph {et~al.}(2015)\citenamefont {Wang},
  \citenamefont {Kumar}, \citenamefont {Jendrzejewski}, \citenamefont {Wilson},
  \citenamefont {Edwards}, \citenamefont {Eckel}, \citenamefont {Campbell},\
  and\ \citenamefont {Clark}}]{Wang_2015}%
  \BibitemOpen
  \bibfield  {author} {\bibinfo {author} {\bibfnamefont {Y.-H.}\ \bibnamefont
  {Wang}}, \bibinfo {author} {\bibfnamefont {A.}~\bibnamefont {Kumar}},
  \bibinfo {author} {\bibfnamefont {F.}~\bibnamefont {Jendrzejewski}}, \bibinfo
  {author} {\bibfnamefont {R.~M.}\ \bibnamefont {Wilson}}, \bibinfo {author}
  {\bibfnamefont {M.}~\bibnamefont {Edwards}}, \bibinfo {author} {\bibfnamefont
  {S.}~\bibnamefont {Eckel}}, \bibinfo {author} {\bibfnamefont {G.~K.}\
  \bibnamefont {Campbell}},\ and\ \bibinfo {author} {\bibfnamefont {C.~W.}\
  \bibnamefont {Clark}},\ }\bibfield  {title} {\bibinfo {title} {Resonant
  wavepackets and shock waves in an atomtronic {SQUID}},\ }\href
  {https://doi.org/10.1088/1367-2630/17/12/125012} {\bibfield  {journal}
  {\bibinfo  {journal} {New Journal of Physics}\ }\textbf {\bibinfo {volume}
  {17}},\ \bibinfo {pages} {125012} (\bibinfo {year} {2015})}\BibitemShut
  {NoStop}%
\bibitem [{\citenamefont {Gallem\'{\i}}\ \emph {et~al.}(2018)\citenamefont
  {Gallem\'{\i}}, \citenamefont {Pitaevskii}, \citenamefont {Stringari},\ and\
  \citenamefont {Recati}}]{PhysRevA.97.063615}%
  \BibitemOpen
  \bibfield  {author} {\bibinfo {author} {\bibfnamefont {A.}~\bibnamefont
  {Gallem\'{\i}}}, \bibinfo {author} {\bibfnamefont {L.~P.}\ \bibnamefont
  {Pitaevskii}}, \bibinfo {author} {\bibfnamefont {S.}~\bibnamefont
  {Stringari}},\ and\ \bibinfo {author} {\bibfnamefont {A.}~\bibnamefont
  {Recati}},\ }\bibfield  {title} {\bibinfo {title} {Magnetic defects in an
  imbalanced mixture of two {Bose-Einstein} condensates},\ }\href
  {https://doi.org/10.1103/PhysRevA.97.063615} {\bibfield  {journal} {\bibinfo
  {journal} {Phys. Rev. A}\ }\textbf {\bibinfo {volume} {97}},\ \bibinfo
  {pages} {063615} (\bibinfo {year} {2018})}\BibitemShut {NoStop}%
\bibitem [{\citenamefont {Knoop}\ \emph {et~al.}(2011)\citenamefont {Knoop},
  \citenamefont {Schuster}, \citenamefont {Scelle}, \citenamefont {Trautmann},
  \citenamefont {Appmeier}, \citenamefont {Oberthaler}, \citenamefont
  {Tiesinga},\ and\ \citenamefont {Tiemann}}]{PhysRevA.83.042704}%
  \BibitemOpen
  \bibfield  {author} {\bibinfo {author} {\bibfnamefont {S.}~\bibnamefont
  {Knoop}}, \bibinfo {author} {\bibfnamefont {T.}~\bibnamefont {Schuster}},
  \bibinfo {author} {\bibfnamefont {R.}~\bibnamefont {Scelle}}, \bibinfo
  {author} {\bibfnamefont {A.}~\bibnamefont {Trautmann}}, \bibinfo {author}
  {\bibfnamefont {J.}~\bibnamefont {Appmeier}}, \bibinfo {author}
  {\bibfnamefont {M.~K.}\ \bibnamefont {Oberthaler}}, \bibinfo {author}
  {\bibfnamefont {E.}~\bibnamefont {Tiesinga}},\ and\ \bibinfo {author}
  {\bibfnamefont {E.}~\bibnamefont {Tiemann}},\ }\bibfield  {title} {\bibinfo
  {title} {Feshbach spectroscopy and analysis of the interaction potentials of
  ultracold sodium},\ }\href {https://doi.org/10.1103/PhysRevA.83.042704}
  {\bibfield  {journal} {\bibinfo  {journal} {Phys. Rev. A}\ }\textbf {\bibinfo
  {volume} {83}},\ \bibinfo {pages} {042704} (\bibinfo {year}
  {2011})}\BibitemShut {NoStop}%
\bibitem [{\citenamefont {Bao}\ and\ \citenamefont
  {Cai}(2015)}]{doi:10.1137/140979241}%
  \BibitemOpen
  \bibfield  {author} {\bibinfo {author} {\bibfnamefont {W.}~\bibnamefont
  {Bao}}\ and\ \bibinfo {author} {\bibfnamefont {Y.}~\bibnamefont {Cai}},\
  }\bibfield  {title} {\bibinfo {title} {Ground states and dynamics of
  spin-orbit-coupled bose--einstein condensates},\ }\href
  {https://doi.org/10.1137/140979241} {\bibfield  {journal} {\bibinfo
  {journal} {SIAM Journal on Applied Mathematics}\ }\textbf {\bibinfo {volume}
  {75}},\ \bibinfo {pages} {492} (\bibinfo {year} {2015})}\BibitemShut
  {NoStop}%
\end{thebibliography}%
\end{document}